\documentclass[aps, twocolumn, notitlepage, superscriptaddress, prb]{revtex4-1}

\makeatletter
\AtBeginDocument{\let\LS@rot\@undefined}
\makeatother

\usepackage{subfigure}
\usepackage{textcomp}
\usepackage{graphicx}
\usepackage{amssymb}
\usepackage{amsmath}
\usepackage{bm}
\usepackage{amsmath, amsthm, amssymb}
\usepackage{dsfont}
\usepackage[colorlinks=true,linkcolor=blue,urlcolor=blue,citecolor=blue]{hyperref}
\usepackage{multirow}
\usepackage{url}
\usepackage{bbold}
\usepackage{array}

\newcolumntype{M}[1]{>{\centering\arraybackslash}m{#1}}
\newcolumntype{N}{@{}m{0pt}@{}}

\def\ket#1{\mathinner{|{#1}\rangle}}
\def\braket#1{\mathinner{\langle{#1}\rangle}}

 \makeatletter
\newsavebox{\@brx}
\newcommand{\llangle}[1][]{\savebox{\@brx}{\(\m@th{#1\langle}\)}%
  \mathopen{\copy\@brx\kern-0.5\wd\@brx\usebox{\@brx}}}
\newcommand{\rrangle}[1][]{\savebox{\@brx}{\(\m@th{#1\rangle}\)}%
  \mathclose{\copy\@brx\kern-0.5\wd\@brx\usebox{\@brx}}}
\makeatother

\def\beq{\begin{equation}}
\def\eeq{\end{equation}}
\def\bea{\begin{eqnarray}}
\def\eea{\end{eqnarray}}

\usepackage[usenames, dvipsnames]{color}

\usepackage{soul,xcolor}
\setstcolor{green}

\usepackage{pdfpages} 
\makeatletter
\AtBeginDocument{\let\LS@rot\@undefined}
\makeatother

\begin{document}
\title{Operator front broadening in chaotic and integrable quantum chains}
\author{Javier Lopez-Piqueres}
\affiliation{Department of Physics, University of Massachusetts, Amherst, MA 01003, USA}
\author{Brayden Ware}
\affiliation{Department of Physics, University of Massachusetts, Amherst, MA 01003, USA}
\author{Sarang Gopalakrishnan}
\affiliation{Department of Physics, The Pennsylvania State University, University Park, PA 16802, USA}
\affiliation{Department of Physics and Astronomy, CUNY College of Staten Island, Staten Island, NY 10314, USA}
\author{Romain Vasseur}
\affiliation{Department of Physics, University of Massachusetts, Amherst, MA 01003, USA}

\begin{abstract}

Operator spreading under unitary time evolution has attracted a lot of attention recently,  as a way to probe many-body quantum chaos. While quantities such as out-of-time-ordered correlators (OTOC) do distinguish interacting from non-interacting systems, it has remained unclear to what extent they can truly diagnose chaotic {\it vs} integrable dynamics in many-body quantum systems. Here, we analyze operator spreading in generic 1D  many-body quantum systems using a combination of matrix product operator (MPO) and analytical techniques, focusing on the operator {\em right-weight}. First, we show that while small bond dimension MPOs allow one to capture the exponentially-decaying tail of the operator front, in agreement with earlier results, they lead to significant quantitative and qualitative errors for the actual front --- defined by the maximum of the right-weight. We find that while the operator front broadens diffusively in both integrable and chaotic interacting spin chains, the precise shape and scaling of the height of the front in integrable systems is anomalous for all accessible times. We interpret these results using a quasiparticle picture. This provides a sharp, though rather subtle signature of many-body quantum chaos in the operator front. 

\end{abstract}

\maketitle


\section{Introduction}

Understanding the propagation of quantum information in many-body quantum systems has become a central theme of modern condensed matter physics. Quantum information dynamics sheds light on such seemingly unrelated problems, including fault-tolerant quantum computing, the foundations of statistical mechanics~\cite{PhysRevA.43.2046,PhysRevE.50.888,goldstein2010long}, the physics of black holes~\cite{hayden2007black}, or holography~\cite{shenker2014black}. This renewed interest in quantum information quantities was partly sparked by recent experimental developments that explored questions related to thermalization of isolated quantum systems~\cite{trotzky2012probing,gring2012relaxation,schneider2012fermionic,cheneau2012light}, and even characterized thermalization or lack thereof by measuring directly entanglement entropies~\cite{Islam:2015aa,Lukin256,2019arXiv191006024C}. Recently, much progress has been devoted to understanding the spreading of quantum operators under unitary evolution in the Heisenberg picture~\cite{shenker2014black, sekino2008fast,hosur2016chaos,lashkari2013towards,PhysRevX.9.031048,PhysRevX.9.041017,PhysRevB.102.045114}, as a way to characterize the scrambling of information into increasingly non-local observables.  Starting from a local operator ${\cal O}$ at $x=0$, an especially interesting quantity is the operator ``front'' (or ``lightcone'') at a given time $t$, corresponding to the location of the farthest non-identity operators in ${\cal O}(t)$. In lattice quantum systems, the operator front is constrained by the Lieb-Robinson bound~\cite{cmp/1103858407}.   

Conventional linear response functions do not diagnose the operator front, since conventional observables relax locally. For example, in the presence of conserved quantities,  autocorrelation functions spread diffusively while the operator front spreads ballistically~\cite{KimHuse,  KnapScrambling, LuitzScrambling, kvh, PhysRevX.8.031058}. Instead, the dynamics of the operator front can be captured by out-of-time-order commutators (OTOC)~\cite{larkin1969quasiclassical, shenker2014black, maldacena2016bound}
\beq\label{otocdef}
{C}({ x},t) \equiv \frac{1}{2} \langle [V(t), W]^\dagger [V(t), W] \rangle,
\eeq
where $V, W$ are local operators separated by a distance $x$, and  the expectation value is taken in a chosen equilibrium ensemble. Initially, the operators $V$ and $W$ are well separated so $C \approx 0$, but as $V(t)$ spreads in a chaotic system, $C$ approaches an ${\cal O}(1)$ number as the lightcone of $V(t)$ overlaps $W$. The general behavior of OTOCs has been studied in various contexts, ranging from random circuits to many-body localized systems~\cite{fan2017out,PhysRevB.95.060201,chen2017out,von2018operator,nahum2018operator,PhysRevB.97.144304,KnapScrambling,xu2020accessing,PhysRevX.8.031058,gopalakrishnan2018hydrodynamics,kvh}, and a number of proposals and subsequent promising experiments have been carried out in the past few years to measure OTOCs directly~\cite{PhysRevA.94.040302,zhu2016measurement,garttner2017measuring, yao2016interferometric,PhysRevLett.123.090605,2021arXiv210108870M}.

A natural question that arises in this context is whether studying operator spreading may elucidate key differences in the dynamics between integrable and chaotic systems \cite{PhysRevE.75.015202,PhysRevLett.122.250603}. In fact, the main motivation for the line of study of operator spreading and OTOCs  was their promise to probe ``many-body quantum chaos'', which in turn should allow one to distinguish integrable {\it vs} chaotic quantum dynamics. Integrable systems have extensively many conservation laws~\cite{PhysRevLett.96.136801,PhysRevLett.106.217206,PhysRevLett.110.257203,PhysRevLett.113.117202,PhysRevLett.115.157201,ilievski2016quasilocal,vasseur2016nonequilibrium,PhysRevB.89.125101,alba2017entanglement}, as well as stable quasiparticle excitations even at high temperature. It is thus natural to expect the dynamics of operator spreading in these systems to differ from that in chaotic systems. 
Initial attempts to establish such a distinction compared fully chaotic systems, such as random circuits, to \emph{free-fermion} models, or models such as the transverse-field Ising model that can be mapped to free fermions. The spreading of a generic operator in a chaotic system is indeed very different from that of, say, a fermion bilinear: notably, in the latter case, the squared commutator~\eqref{otocdef} \emph{vanishes} at late times inside the front, instead of saturating. However, it was realized that there are natural operators in the transverse-field Ising model---such as the local order parameter---for which the squared commutator saturates behind the front. A more refined distinction was then sought, based on the idea that fronts in chaotic systems seem to broaden diffusively in time, as one can prove for random circuits~\cite{von2018operator,nahum2018operator}; meanwhile,
non-interacting systems have operator fronts that broaden sub-diffusively~\cite{Platini2005, PhysRevB.97.081111,2018arXiv180305902K,xu2020accessing,PhysRevB.97.144304,PhysRevB.96.220302} as $t^{1/3}$. The behavior of the OTOC behind the front in free fermion systems also shows a pattern of oscillations~\cite{PhysRevB.97.144304}, in sharp contrast with generic chaotic systems where the OTOC approaches a universal constant behind the front.  

However, {\em interacting} integrable quantum systems also have an operator front that spreads ballistically and broadens diffusively~\cite{gopalakrishnan2018hydrodynamics}. Although the mechanisms for this behavior are very different than in quantum chaotic systems (in interacting integrable systems, diffusion is due to random time delays as a result of collisions between different 	quasiparticles~\cite{gopalakrishnan2018hydrodynamics,de2019diffusion}) no qualitative distinctions between the operator fronts of interacting integrable systems and chaotic systems have yet been identified.
Even in integrable cellular automata, a class of exceedingly simple interacting integrable systems, operator spreading probes are hard to distinguish from the chaotic case~\cite{Gopalakrishnan_2018,PhysRevB.98.060302}.

In this work, we identify ways in which the operator front differs in integrable {\it vs} chaotic many-body quantum systems. To carry out the analysis we focus on the \textit{right-weight} of a given operator~\cite{von2018operator,nahum2018operator}, which measures the spreading of an initially local operator under Heisenberg time evolution propagating to the right of a 1D system. This quantity has the advantage of being peaked at the operator front compared to the OTOC. (For unitary random circuits, the right-weight is simply related to the OTOC by a spatial derivative~\cite{von2018operator,nahum2018operator}, but the two quantities are distinct in general.) We study the right-weight using matrix product operators (MPO) techniques~\cite{xu2020accessing,PhysRevB.100.104303}. One of our key observations is that while small bond dimension MPOs do allow one to capture the exponentially-decaying tail of the front~\cite{xu2020accessing,PhysRevB.100.104303} (and describe the front exactly in the case of dual-unitary quantum circuits~\cite{PhysRevResearch.2.033032, PhysRevLett.123.210601, bertini2020operator, PhysRevB.100.064309}), they lead to significant quantitative and qualitative errors for the actual front (defined by the maximum of the right-weight).
This feature is especially obvious when considering the right-weight compared to OTOCs. Truncation errors are actually fairly dramatic: for chaotic systems, we find that the operator front stops moving at finite time, or even disappears at small bond dimensions. For integrable systems, small bond dimensions MPOs lead to operator fronts broadening subdiffusively (close to $\sim t^{1/3}$ as in non-interacting systems)~\cite{xu2020accessing}, while the front is expected to broaden diffusively on general grounds~\cite{gopalakrishnan2018hydrodynamics}. We show that this discrepancy is resolved by considering larger bond dimensions, and confirm numerically that the operator front does broaden diffusively in interacting integrable quantum systems. 

Armed with these results, we analyze numerically how the operator front broadens in integrable systems. We find that contrary to chaotic systems where the diffusive front broadening is characterized by a Gaussian function, the operator front in integrable systems scales anomalously for all accessible times. In particular, we find that the height of the operator front in such systems decays as $t^{-3/4}$, instead of $t^{-1/2}$ in the non-integrable case. We explain these results using a quasiparticle picture of operator spreading, and compute the universal scaling function characterizing the front broadening in integrable systems. 

The organization of the paper is as follows. In Sec. \ref{sec: background} we introduce the main object under study in this work, the right-weight of a given operator which will allow us to analyze operator front-broadening in the rest of the manuscript. We also briefly discuss some basics of the \textit{time-evolving block decimation} (TEBD) numerical algorithm adapted to the study of operator spreading. We also discuss truncation errors due to the finite bond dimension of matrix product operators. 
We present the results of operator front-broadening in Sec. \ref{sec: integrable} for integrable systems and in Sec. \ref{sec: non-integrable} for non-integrable systems. We conclude in Sec. \ref{sec: conclusion} and give an outlook for future work.

 \begin{figure*}[t!]
\centering
\includegraphics[width=\linewidth,clip]{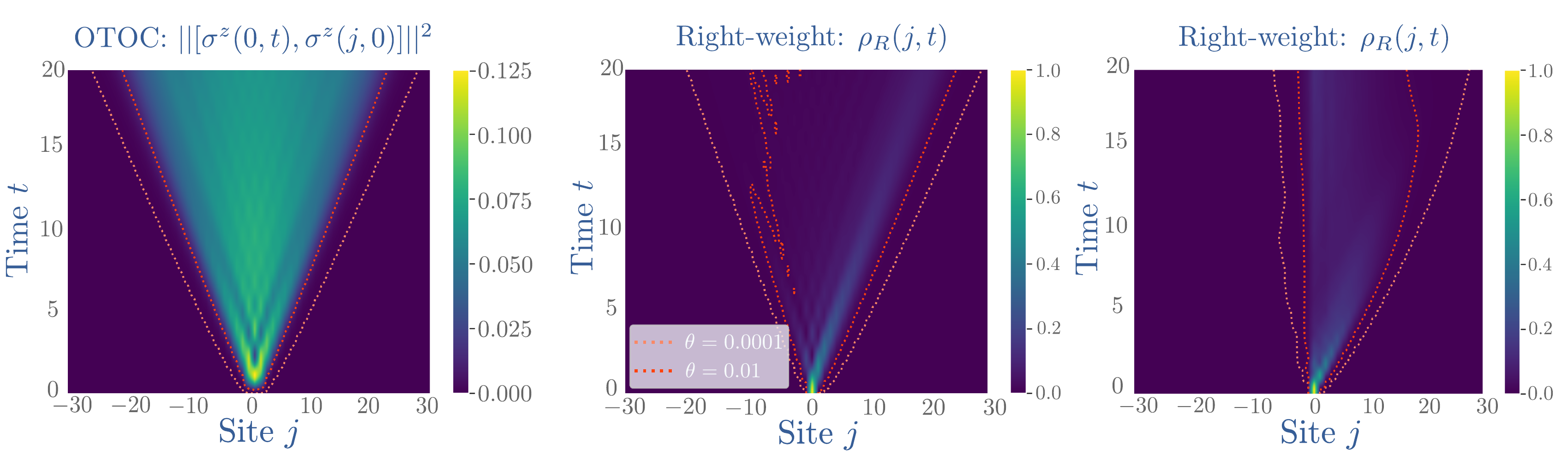}
\caption{\textbf{Operator spreading in integrable and chaotic spin chains} starting from $\mathcal{O}_0=\sigma_0^z$. Left panel: OTOC spatio-temporal profile in the integrable XXZ model for $\Delta=0.5$ and $\mathcal{V}_x=\sigma_x^z$. Middle panel: right-weight profile still in the integrable XXZ model with $\Delta=0.5$. Right panel: right-weight profile in the non-integrable \textit{transverse-field Ising} model with $h_z=0.9045$ and $h_x=0.8090$. The dashed lines are contour lines following a given threshold $\theta$. Data obtained with bond dimension $\chi_{\rm MAX}=128$. For integrable models, both the OTOC and the right-weight behave in a reasonable way, despite the relatively small bond dimension, but the front broadens subdiffusively because of truncation errors. In the chaotic case (right panel), the operator front disappears at finite time because of truncation errors. }
\label{Fig: right-weight heatmap}
\end{figure*}

\section{Operator Right-weight, matrix product operators and truncation errors} \label{sec: background}

In this section, we introduce our main quantity of interest, the operator {\em right-weight}, and explain how it can be computed numerically using MPOs. We also address the effects of truncation errors. 

\subsection{Operator right-weight}
Consider the spreading of an initially local operator $\mathcal{O}_0\equiv \mathcal{O}(x=0)$ under Heisenberg time evolution. Under time evolution, this operator will grow into a more complicated one $\mathcal{O}(t)=[U(t)]^\dagger \mathcal{O}_0U(t)$ being a superposition of many \textit{strings} made of products of non-trivial local operators. A way to characterize the complexity of this object is by means of the OTOC. Consider another local operator at site $x$, $\mathcal{V}_x$. The OTOC is defined as the squared of the commutator between these two operators $\mathcal{C}(x,t)\equiv ||[\mathcal{O}(t),\mathcal{V}_x]||^2=2\left(1-\text{Re}(\text{tr}[\mathcal{O}(t)^\dagger \mathcal{V}_x^\dagger \mathcal{O}(t)\mathcal{V}_x])\right)$. The shape of the OTOC shares universal features across generic systems including ballistic spreading of the wavefront, a rapid growth ahead of the wavefront and saturation behind the wavefront at late times. These features are showcased in Fig.~\ref{Fig: right-weight heatmap} for the integrable XXZ model. 

To characterize the size of an initially local operator $\mathcal{O}_0$ under Heisenberg evolution consider instead the decomposition
\begin{equation} \label{Eq: o(t)}
\mathcal{O}(t)=\sum_{\mathcal{S}}a_{\mathcal{S}}(t)\mathcal{S},
\end{equation}
where the sum above goes over all possible string operators (Pauli strings in the case of spin-1/2 operators). A complete understanding of operator spreading can be captured by the set of coefficients $\{a_{\mathcal{S}}(t)\}$, a task which is out of reach. Instead, we are interested in coarse-grained quantities relating these coefficients. One such quantity is the \textit{right-weight}. For a given operator $\mathcal{O}(t)$ it reads \cite{nahum2018operator, von2018operator}: 
\begin{equation} \label{Eq: rweight}
\rho_R(j,t)=\sum_{\substack{\text{strings w/} \\ \text{rightmost} \\ \text{non-identity} \\ \text{on site j}}}|a_{\mathcal{S}}(t)|^2.
\end{equation}
The coefficients $a_{\mathcal{S}}(t)$ appearing in the expression can be obtained exploiting the fact that these strings form an orthogonal basis in a Hilbert space of dimension $\mathcal{D}^2$: $a_\mathcal{S}(t)=\text{tr}\left[\mathcal{S}^\dagger \mathcal{O}(t)\right]/\mathcal{D}$. (Here $\mathcal{D}$ is the Hilbert space dimension of states; i.e. $\mathcal{D}=2^L$ for spin-$1/2$ chains). We require the initial operator to be normalized, i.e. $\text{tr}\left[\mathcal{O}_0^\dagger \mathcal{O}_0\right]/\mathcal{D}=1$, which implies (using unitarity) the sum rule: $\sum_{\mathcal{S}}|a_{\mathcal{S}}(t)|^2=1$. Note that by construction we also have $\sum_j \rho_R(j,t) =1$. This conservation law has important consequences for the ``hydrodynamics'' of operator spreading in both integrable and non-integrable systems. On general grounds, we expect the associated current to behave as $j = v_B \rho_R - D \partial_x \rho_R + \dots$, where $v_B$ is the butterfly velocity characterizing the speed of the ballistically moving operator front, $D$ is a diffusion constant that sets the generic diffusive broadening of the front, and the dots represent non-linear and higher-derivative terms. In what follows we shall focus on spin-$1/2$ chains, both integrable and chaotic.

\subsection{Matrix product operators}\label{Sec: TEBD}

In order to measure the right-weight numerically, we use matrix product operator (MPO) techniques. For this purpose, we express Eq. (\ref{Eq: o(t)}) as a state in the Hilbert space of operators as is routinely done in the context of time evolution of MPOs \cite{paeckel2019time}, so that $\mathcal{O}(t)\rightarrow |\mathcal{O}(t)\rrangle\equiv \sum_{\mathcal{S}}a_\mathcal{S}(t)|\mathcal{S}\rrangle$. To evaluate the right-weight as a correlator, we introduce the projector onto the identity acting on site $x$, $\mathcal{P}_{\mathbb{1},x}$ (i.e. $\mathcal{P}_{\mathbb{1},x}\equiv |\mathbb{1}\rrangle _x\llangle \mathbb{1}|_x$), where $|\mathbb{1}\rrangle\equiv\otimes_{x=1}^L\left(\ket{00}_x+\ket{11}_x\right) / \sqrt{2}$.  We reserve odd entries of any MPS in this newly enlarged Hilbert space for the \textit{physical} sites, and the even sites for the \textit{ancilla} sites\cite{paeckel2019time}. It is then straightforward to show that the right-weight can be computed as follows
\begin{equation}
\rho_R(x,t)=\frac{\partial}{\partial x}\llangle \mathcal{O}(t)|\prod_{x'\geq x}\mathcal{P}_{\mathbb{1},x'}|\mathcal{O}(t)\rrangle, \label{rightweightMPO}
\end{equation}
where $\partial_x$ should be interpreted as a discrete spatial derivative. 

We compute the right-hand side of eq.~\eqref{rightweightMPO} using the TEBD algorithm~\cite{PhysRevLett.91.147902, PhysRevLett.93.040502} applied to matrix product operators. We denote the maximum bond dimension as $\chi_{\rm MAX}$. Time evolution is implemented directly in operator space as $|\mathcal{O}(t) \rrangle = e^{-it\mathcal{L}}|\mathcal{O}_0 \rrangle$ where $\mathcal{L}\equiv-H\otimes \mathbb{1}+\mathbb{1}\otimes H^T$, where the Kronecker product here is used to distinguish physical from ancilla space.  In this language, standard two-point correlation functions can be computed as simple overlaps between states in this doubled Hilbert space. 

In our numerical simulations, unless otherwise stated, we will be considering a system size of $L=401$ sites, a fourth order Trotter decomposition of step size $dt=0.1$ and a cutoff error of $\epsilon=10^{-10}$. The system size was chosen so that the right-weight front never reaches the boundary of the system within the time scale of interest, which is $t_{\rm MAX}\sim \mathcal{O}(10^2)$. These simulations are carried out using the C++ iTensor library~\cite{fishman2020itensor}.

\subsection{Operator front and truncation errors}

In the remainder of this paper, we will use this MPO approach to compute the right-weight in various interacting chaotic and integrable spin $1/2$ chains. Before we address specific features of operator spreading in those different classes of systems, we address here the dramatic effects of truncation errors in the MPO approach. Representative plots of the right-weight and of OTOCs are shown in Fig.~\ref{Fig: right-weight heatmap}, for both a chaotic Ising chain, and for the integrable XXZ spin chain, using a finite bond dimension $\chi_{\rm MAX} =128$.

For integrable chains, both the OTOC and the right-weight behave as expected, despite the finite bond dimension. However, as we will show below, some qualitative details end up being affected by the truncation errors. In particular, for finite bond dimension, we will see that the front broadens subdiffusively as $t^{\alpha}$ with $\alpha \approx 1/3$ for small bond dimensions, while we recover $\alpha=1/2$ as $\chi_{\rm MAX} \to \infty$. This explains the apparent $t^{1/3}$ broadening in the integrable Heisenberg chain observed in Ref.~\onlinecite{xu2020accessing} using small bond dimension MPOs. 

The effects of truncation errors on the operator front in chaotic chains are much more dramatic. As shown in the right panel of Fig.~\ref{Fig: right-weight heatmap}, the operator front (defined as the maximum of the right-weight, moving at the butterfly velocity $v_B$) fades away and disappears at short times. We will show below that this unphysical feature is entirely due to truncation errors, and can be deferred to longer times by increasing the bond dimension. Thus large bond dimensions are absolutely essential to describe the operator front correctly. 
In contrast,  bond dimensions as low as $\chi_{\rm MAX} =4$ can be enough to capture the exponentially-decaying tails of the operator front, as noted in Refs.~\onlinecite{xu2020accessing,PhysRevB.100.104303}. Our results are also consistent with  contour lines for the OTOC being less than a given threshold $\epsilon$ being less sensitive to truncation errors for small $\epsilon$  (see dashed line in Fig.~\ref{Fig: right-weight heatmap}). 
However, as we show here, the small-$\epsilon$ contours \emph{outside} the front are an unreliable guide to the location of the front itself (i.e., the maximum of the right-weight). In the case of integrable systems, using those tails to analyze the front broadening gives rise to incorrect results for low bond dimensions. 
%
%
In the following, we will carefully analyze the convergence of our results with respect to bond dimension; for practical purposes we restrict ourselves to maximal bond dimensions less than $\chi_{\rm MAX} =512$ in most cases to access long enough times.

\section{Operator front in integrable systems}\label{sec: integrable}

Armed with this numerical tool, we analyze the operator front in integrable quantum systems. As in chaotic systems, we expect a ballistically moving front, broadening as $t^{1/3}$ in free systems~\cite{Platini2005, PhysRevB.97.081111,2018arXiv180305902K,xu2020accessing,PhysRevB.97.144304,PhysRevB.96.220302}, and $t^{1/2}$ in interacting integrable systems~\cite{gopalakrishnan2018hydrodynamics}. In integrable systems, we expect the operator front to follow the fastest quasiparticle. For interacting integrable systems, quasiparticles behave as biased random walkers due to their random collisions with other quasiparticles~\cite{PhysRevB.98.060302, gopalakrishnan2018hydrodynamics, PhysRevLett.121.160603, Gopalakrishnan_2018,de2019diffusion}. In the following, we will confirm those predictions numerically, but also identify a key difference with chaotic systems. As we will show, the quasiparticle picture suggests that the peak height of the front decays anomalously as $t^{-3/4}$, at least at intermediate times, and scales with a non-Gaussian universal function that we compute exactly. 

\subsection{Free fermions}

Before turning to {\em interacting} integrable quantum systems, we briefly recall how operators spread in spin chains dual to free fermions, following Ref.~\onlinecite{PhysRevB.97.144304}. For concreteness, we focus on the XX spin chain with Hamiltonian 
\begin{equation}
H=J\sum_j S_j^xS_{j+1}^x+S_j^yS_{j+1}^y,
\end{equation}
where $S^\alpha_j$ are spin-1/2 operators acting on site $j$, and $J=1$ in the following. Let us consider the spreading of the Pauli operator $\sigma^z$ of the $XX$ model initially at site $0$, that is $\mathcal{O}_0=\sigma_{0}^z$. Since this Hamiltonian is Jordan-Wigner dual to free fermions, this reduces the possible Pauli strings participating in $\mathcal{O}_0(t)$. 
Out of the $4^L$ possible Pauli strings, only $L^2$ Pauli strings will contribute here. Indeed only the operators $\sigma_i^+\left(\prod_{i<l<j}\sigma_l^z\right)\sigma_j^-$, $\sigma_j^z$, for general $i,j$, will contribute, as those are the only spin operators that map to quadratic fermions under a Jordan-Wigner transformation.  Thus, using this fact and (\ref{Eq: rweight}) the right-weight of $\mathcal{O}_0$ takes the form 
\begin{align} \label{eq_rweight_free}
\begin{split}
\rho_R(j,t)=|a_{\sigma_j^z}(t)|+\sum_{i<j}|a_{\sigma_i^+(\prod_{i<l<j}\sigma_l^z)\sigma_j^-}(t)|^2\\ 
+|a_{\sigma_i^-(\prod_{i<l<j}\sigma_l^z)\sigma_j^+}(t)|^2. 
\end{split}
\end{align}
The correlators appearing in (\ref{eq_rweight_free}) can be evaluated straightforwardly by mapping spin operators to spinless fermions to yield
\begin{equation}
\rho_R(x,t)=[J_x(t)]^4+2[J_x(t)]^2\sum_{y<x}[J_{y}(t)]^2, 
\end{equation}
were  $J_x(t)\equiv 1/2\pi\int_{-\pi}^\pi \text{e}^{-i(kx+\cos(k)t)}dk$ are Bessel functions of the first kind. As $t$ and $x$ become large, this yields the following scaling form for the right-weight \cite{PhysRevE.69.066103}
\begin{equation}
\rho_R(x,t) \sim  \frac{1}{t^{2/3}}F\left(\frac{x-t}{t^{1/3}}\right),
\end{equation}
where the butterfly velocity is $v_B=1$, and $F$ is some universal scaling function. This establishes that the operator front broadens subdiffusively as $t^{1/3}$ in free fermion systems. 

\subsection{Interacting integrable spin chains} \label{Sec: integrable-interacting}

\begin{figure}
\centering
\includegraphics[width=\linewidth,clip]{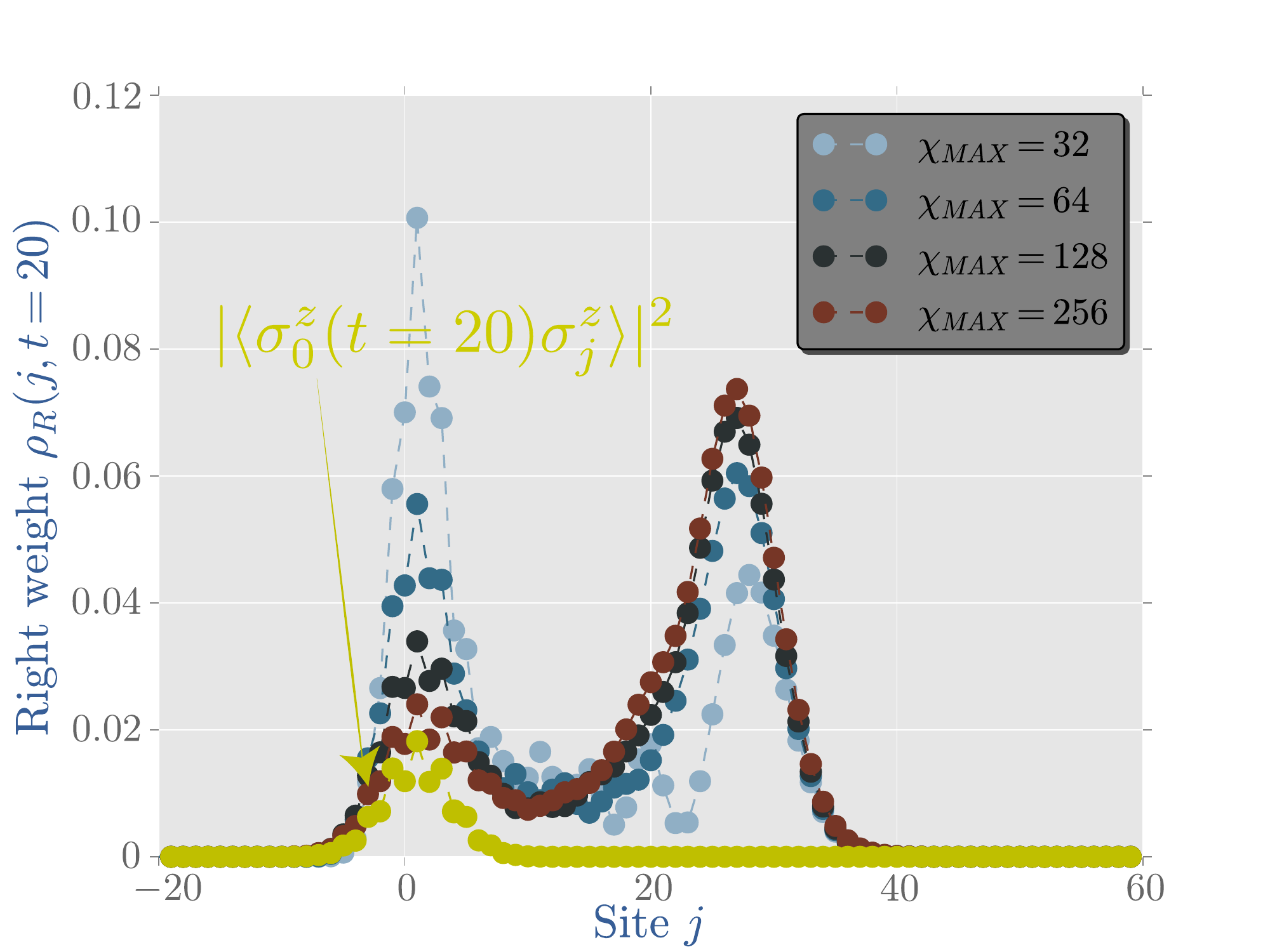}
\caption{\textbf{Right-weight spatial profile} at $t=20$ in the XXZ spin chain with $\Delta=5$, and $\mathcal{O}_0=\sigma^z_{0}$ for various maximum bond dimensions. The yellow data points correspond to the squared correlator $|\braket{\sigma_{0}^z(t=20)\sigma_j^z}|^2$ at infinite temperature $\beta = 0$, which lower bounds the right-weight. }
\label{Fig: right_weight_prof_Delta_5}
\end{figure}

\begin{figure}
\centering
\includegraphics[width=\linewidth,clip]{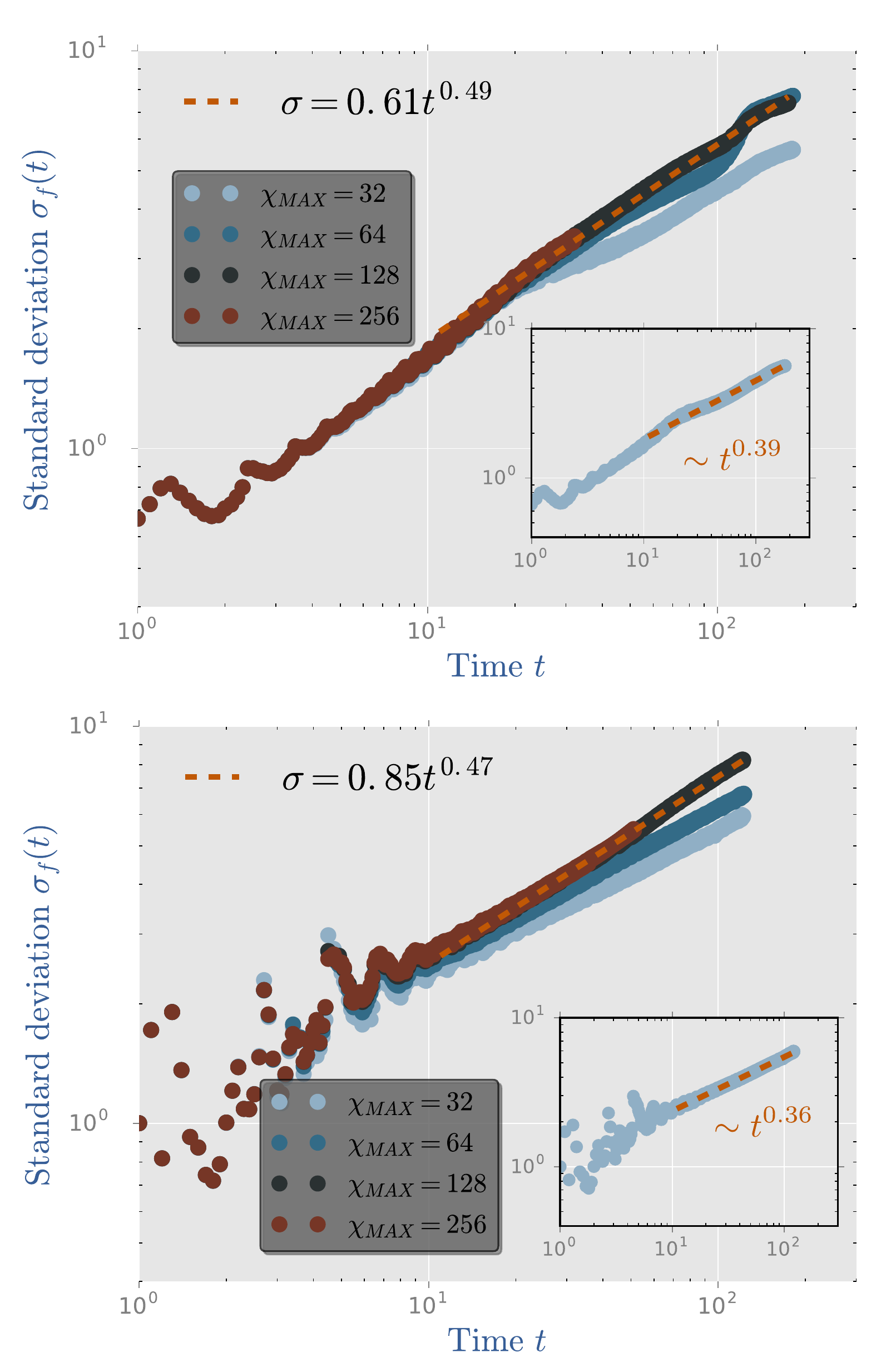}
\caption{\textbf{Standard deviation of the front of the right-weight versus time} for various maximum bond dimensions plus a linear fit over the data for $\chi_{\rm MAX}=128$ showing approximately diffusive spreading. Top panel: $\Delta=1/2$ and $\mathcal{O}_0=\sigma^z_{0}$. Bottom panel: $\Delta=5$ and $\mathcal{O}_0=\sigma^x_{0}$. Insets in both panels depict the standard deviation of the front for a small bond dimension $\chi_{\rm MAX}=32$, showing that truncation errors lead to an operator front that broadens subdiffusively with an exponent close to $1/3$.}
\label{Fig: spreading_Deltas}
\end{figure}

We now turn our attention to operator spreading in {\em interacting} integrable systems. Our model of interest will be the paradigmatic spin-$1/2$ XXZ Hamiltonian
\begin{equation}
H=J\sum_j S_j^xS_{j+1}^x+S_j^yS_{j+1}^y+\Delta S_j^zS_{j+1}^z,
\end{equation}
In what follows, we set $J=1$. This model is integrable and in this sense ``exactly solvable'', though quantities such as OTOC or the right-weight are analytically out of reach, and have to be computed numerically. 

We analyze numerically the right-weight $\rho_R(j,t)$ for various values of the anisotropy $\Delta$. In what follows, we mostly focus on the initial operator $\sigma^z_{0}$, but we will also consider other operators.  A typical plot of the right-weight at a given time (here $t=20$), for $\Delta =5$ is shown in Fig.~\ref{Fig: right_weight_prof_Delta_5}, for different maximum bond dimensions. A few key features are worth noting. First, as already anticipated above, the operator front -- corresponding to the right-moving peak in the right-weight -- clearly requires large bond dimensions to be captured accurately. Second, the right-weight also shows a diffusively-spreading lump near the origin, lagging behind the operator front. This is a signature of the diffusive spin transport in this model~\cite{ljubotina2017spin}: the right-weight is lower bounded by the square of the infinite-temperature spin autocorrelation function $|\braket{\sigma_{j}^z(t)\sigma_0^z}|^2$, which is known to behave diffusively in the XXZ spin chain for $\Delta >1$~\cite{ljubotina2017spin}. The effects of $U(1)$ conservation laws on operator spreading in chaotic systems was studied in Refs.~\onlinecite{kvh,PhysRevX.8.031058}, and is qualitatively similar in the XXZ spin chain with $\Delta>1$, as finite-temperature spin transport is diffusive in this regime \cite{de2019diffusion} (see also Ref. ~\onlinecite{bertini2021finite}). In contrast, when $\Delta<1$, spin transport in this system is known to be ballistic, and we do not observe a lump of right-weight near the origin (Fig.~\ref{Fig: right-weight heatmap}, left panel). The right-weight in this regime is still nontrivially lower-bounded by the dynamical correlation function; however, in this case the dynamical correlation function scales as $1/t$ all the way out to the light cone, so one does not expect a visible lump near the origin.

In integrable systems, we expect the operator front to coincide with the speed of the fastest quasiparticle~\cite{gopalakrishnan2018hydrodynamics}. As a result, the butterfly velocity should depend on the density of all other quasiparticles, and thermal fluctuations naturally give rise to diffusive broadening of the front.  
To check this numerically, we compute the width of the operator front for an initially local operator as a function of time.  By computing the standard deviation of the front of the right-weight for both $\Delta=1/2$ and $\Delta=5$ (Fig.~\ref{Fig: spreading_Deltas}), we find that the operator front does broaden as $\sigma_f \sim t^{a}$ with $a\sim 0.5$. 
As anticipated above, our results  show that large bond dimensions are required to capture this diffusive broadening of the front (with bond dimensions larger than $\chi_{\rm MAX}\sim 10^2$). Below that threshold, the results do not converge at intermediate to large times, and we find instead some apparent subdiffusive front broadening (see insets in Fig. \ref{Fig: spreading_Deltas}). 

An intuitive way to understand why one cannot restrict to low maximum bond dimension to study the \textit{entire} operator front is to realize that finite bond dimension \textit{truncations} are a non-local operation: while the tail is well-captured by a low maximum bond dimension (since this lies outside the lightcone, where the MPO is represented by lightly entangled blocks) at short enough times, the width of the front is affected in a non-trivial way because of truncations deep in the light cone (see Fig.~\ref{Fig: right_weight_prof_Delta_5}). 

\begin{table}
  \centering
  \begin{tabular}{|M{1.8cm}||M{1.8cm}|M{2.2cm}|M{1.8cm}|N}
    \cline{2-4}
    \multicolumn{1}{c|}{} & Free & Integrable  & Chaotic \\ \hline \hline 
    $\alpha$ & $1/3$   & $1/2$   & $1/2$ & \\[10pt] \hline
    $\beta$ & $2/3$   & $3/4$   & $1/2$ & \\[10pt] \hline
  \end{tabular}
\caption{\textbf{Scaling exponents for generic operator fronts in quantum spin chains} (for intermediate, accessible time scales).  The width of the front scales as $w(t)\sim t^{\alpha}$, while the height scales as $h(t)\sim t^{-\beta}$. 
}
\label{tab: table}  
\end{table} 
\begin{figure}
\centering
\includegraphics[width=\linewidth,clip]{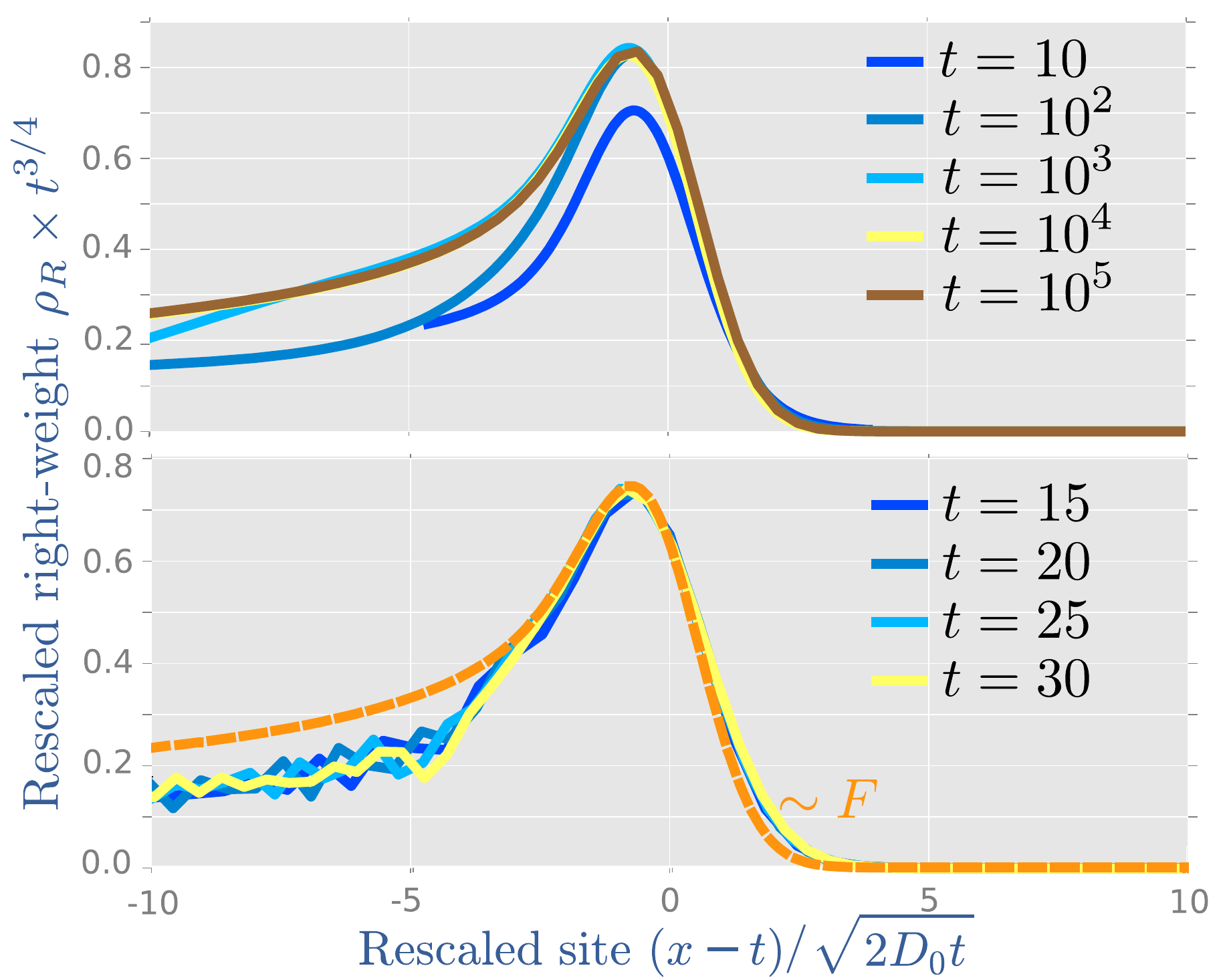}
 \caption{\textbf{Scaling of the front} for the XXZ spin chain with $\Delta=1/2$ and $\mathcal{O}_0=\sigma^z_{0}$. Top panel: collapse of right-weight from the model~\eqref{Eq: toy} for different times and asymptotic form $F(u)$. Here $\omega_\lambda$ was chosen to be Gaussian, though its precise form does not matter. 
   Bottom panel: collapse of right-weight from TEBD at short times.  }
\label{Fig: Bethe vs TEBD}
\end{figure}

\subsection{Scaling of the front and quasiparticle picture of operator spreading}

At the moment, there is no theory for computing quantities like the right-weight (or the OTOCs) in interacting integrable systems. However, it is natural to expect that operator spreading should be captured by the quasiparticles of the underlying integrable model, similar to the quasiparticle picture of entanglement spreading~\cite{PhysRevLett.96.136801, calabrese2007entanglement, fagotti2008evolution, alba2017entanglement, alba2017renyi, alba2018entanglement, alba2020diffusion}. 
Thermodynamics and hydrodynamics in integrable systems can entirely be understood in terms of quasiparticles. This is the basis of the recent framework of generalized hydrodynamics (GHD)~\cite{PhysRevLett.117.207201, castro2016emergent, doyon2019lecture}. Within a given  (generalized) equilibrium state, quasiparticles with quantum number $\lambda$ (called rapidity) move ballistically with a velocity $v_{\rm \lambda}$, with an associated diagonal diffusion constant $D_{\rm \lambda}$ due to random collisions with other quasiparticles in the thermal background. 
Both $v_{\rm \lambda}$ and  $D_{\rm \lambda}$  can be computed analytically in a given generalized equilibrium state. These quasiparticles are known control to transport properties and entanglement scaling, so it is natural to expect them to control operator spreading as well. Let us assume phenomenologically that the right-weight couples to quasiparticles  propagating from the position of the initial operator in a featureless (infinite temperature) background, with an unknown weight $\omega_\lambda$ (normalized so that $\int d \lambda \omega_\lambda =1$). This means that we expect the right-weight to be given by
\begin{equation} \label{Eq: toy}
\rho_R(x,t) \sim \int d\lambda \omega_\lambda \frac{1}{\sqrt{4\pi D_\lambda t}} e^{-\frac{(x-v_\lambda t)^2}{4D_\lambda t}}.
\end{equation}
The weight $\omega_\lambda$ is an unknown function in general. On general grounds, we expect the operator front to be described by the fastest quasiparticle excitation in the system~\cite{gopalakrishnan2018hydrodynamics}. This would correspond to $\omega_\lambda = \delta (\lambda - \lambda_0)$, with $\lambda_0$ the rapidity corresponding to the fastest quasiparticle, and $\rho_R(x,t) = \frac{1}{\sqrt{4\pi D_0 t}} e^{-\frac{(x-v_B t)^2}{4D_0 t}}$, with $D_0 = D_{\lambda_0}$, $v_B = v_{\lambda_0}$. This would be a Gaussian front, as in 1d chaotic systems~\cite{nahum2018operator,von2018operator}(in particular the height of the front should decay as $\sim t^{-1/2}$). Our numerical data is however not consistent with this picture for the times we can access: (1) We find numerically that the speed of the front is slightly lower than $v_{\lambda_0}$, (2) The diffusion constant associated with the diffusive broadening of the front in Fig.~\ref{Fig: spreading_Deltas} does not coincide with the GHD predictions for $D_0$, (3) The operator front observed numerically is clearly non-Gaussian (Fig.~\ref{Fig: Bethe vs TEBD}), and in particular, its height decays as $\sim t^{-3/4}$ (instead of $\sim t^{-1/2}$). 

\begin{figure}
\centering
\includegraphics[width=\linewidth,clip]{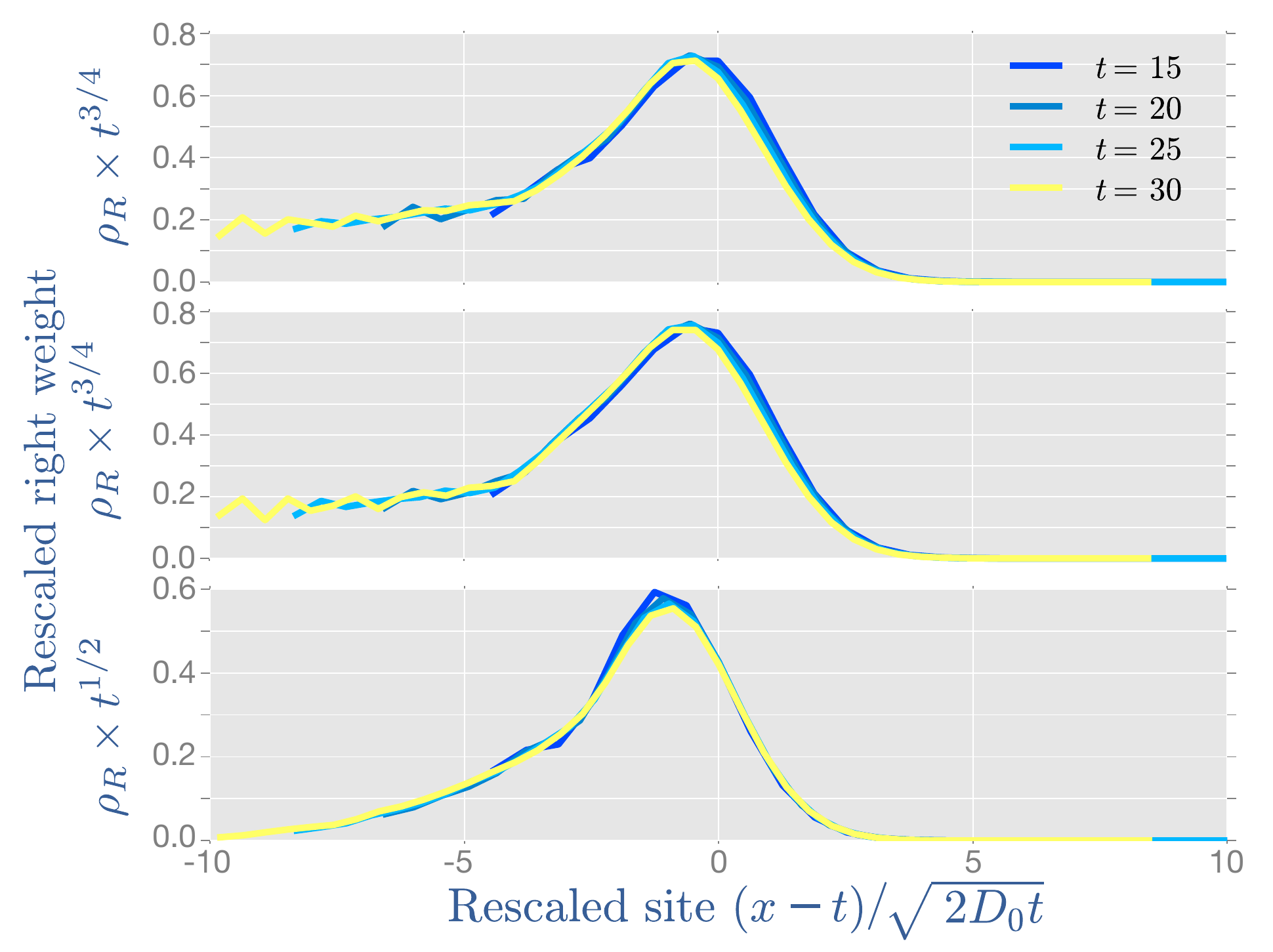}
\caption{\textbf{Collapse of the right-weight for various operators $\mathcal{O}_0$ for the XXZ model with $\Delta=1/2$.} Top panel: $\mathcal{O}_0$ given by the energy density on site 0. Middle panel: local charge of a non-conserved operator that couples to conserved charges $\mathcal{O}_0=\sigma^+_{0}\sigma^-_{1}+\text{h.c.}$. Bottom panel: local charge of a non-conserved operator that does not couple to any conserved charge $\mathcal{O}_0=\sigma^x_{0}$. In this last case, we find that the height of the front scales as $t^{-1/2}$. }
\label{Fig: right-weight collapse ops}
\end{figure}

All those observations indicate that, at least for times accessible within TEBD, the right-weight couples more generically to a continuum of quasiparticles with rapidity near $\lambda_0$. In fact, eq.~\eqref{Eq: toy} predicts a universal form for the operator front as long as $\omega_\lambda$ is non-zero within a finite neighborhood of $\lambda_0$.  The asymptotic behavior of~\eqref{Eq: toy} at long times can then be obtained through a saddle point analysis. Expanding all quantities near the front, we have $v_\lambda = v_B - w (\lambda - \lambda_0)^2 + \dots $, $D_\lambda = D_0 +\dots$, and  $\omega_\lambda = \omega_0 + \dots$, where $w>0$ since by assumption $v_B$ is the maximum velocity.  Plugging these expressions into eq.~\eqref{Eq: toy} and changing variables, we find that 
\begin{equation} \label{EqScaling}
\rho_R(x,t)\sim t^{-3/4}F\left( u \right),
\end{equation}
where $u \equiv (x-v_Bt)/\sqrt{2D_0t}$, with the {\em universal} scaling function
\begin{equation} \label{eqScalingFunction}
F(u)=\int \frac{d\eta}{\sqrt{2\pi}} {\rm e}^{-\frac{1}{2}(u+\eta^2)^2}.
\end{equation}
This intermediate-time scaling form is one of our main results. It is entirely independent of the weight $\omega_\lambda$, as long as the right-weight couples to a continuum of quasiparticles with rapidity near $\lambda = \lambda_0$. The height of the operator front decays as $t^{-3/4}$ rather than decaying as $t^{-1/2}$ as in chaotic systems, with the associated non-Gaussian scaling function~\eqref{eqScalingFunction}. In particular,  we have $ F(u) \sim 1/\sqrt{|u|}$ as $u \to -\infty$, indicating a fat tail behind the front that scales as $t^{-3/4}\times u^{-1/2} \sim \frac{1}{\sqrt{t}}\frac{1}{\sqrt{{v_Bt-x}}}$.

As shown in Fig.~\eqref{Fig: Bethe vs TEBD}, eq~\eqref{Eq: toy} approaches the scaling form~\eqref{EqScaling} at long times only ($t \sim 10^4$) for generic functions $\omega_\lambda$, making it challenging to observe numerically. However, we find that our TEBD data collapses very well against the scaling~\eqref{EqScaling}, even though the resulting collapse is not fully converged to the scaling function~\eqref{eqScalingFunction} at those times (Fig.~\eqref{Fig: Bethe vs TEBD}). Our TEBD data very clearly indicates a non-Gaussian front, with the height decaying with an exponent consistent with $t^{-3/4}$.

In Fig. \ref{Fig: right-weight collapse ops} we show results of the right-weight for various choices of initial operators in the XXZ spin chain.   Operators corresponding to conserved charges, such as energy, are expected to have a right-weight that scales as in (\ref{EqScaling}). Other operators such as $\mathcal{O}_0=\sigma_0^+\sigma_1^- +\text{h.c.}$  are not conserved, but do couple to hydrodynamic modes (in this case energy), and thus are expected to scale as in (\ref{EqScaling}) as well.  To see this note that one may introduce the projector onto hydrodynamic modes $\mathcal{P}=\sum_{i,j}|I_i\rrangle C_{ij}^{-1} \llangle I_j|$,  where the sum goes over all pairs of conserved charges, $\{| I_i \rrangle \}$ is the set of all conserved charges in vector form (using the notation from Sec. \ref{sec: background}), and $C_{ij}=\llangle I_i | I_j\rrangle \equiv 2^{-L} \text{tr}(I_i I_j)$.  Thus any operator $\mathcal{O}$ with $\mathcal{P}|\mathcal{O}\rrangle \neq 0$ is expected to have a corresponding right-weight scaling as in (\ref{EqScaling}) (at least for intermediate times). In comparison,  the last panel in Fig. \ref{Fig: right-weight collapse ops} shows the right-weight of the operator $\sigma^x$; this operator manifestly does not couple to any hydrodynamic modes as it breaks the $U(1)$ symmetry. The behavior of this operator is quite unlike that described above: it has a Gaussian front  that closely resembles what one would see in a chaotic system. In particular, the height of the front scales down as $t^{-1/2}$ rather than as $t^{-3/4}$. The anomalous scaling observed in Figs. \ref{Fig: Bethe vs TEBD}-\ref{Fig: right-weight collapse ops} is stable against increasing bond dimension (data shown for $\chi_{\rm MAX}=256$). In the Appendix we show the scaling collapse for a different conserved operator (in this case $S_z$) for two different bond dimensions, providing numerical evidence that the anomalous scaling is not the result of finite bond dimension.

Our results do not settle the asymptotic late-time behavior of the right-weight in integrable systems. It seems plausible that for \emph{generic} operators there will be some non-hydrodynamic piece (that does not couple to single quasiparticles) in addition to the hydrodynamic piece---we have no reason to expect that the coupling to single quasiparticles exhausts the operator weight. Assuming there is some such non-hydrodynamic piece, the $t^{-1/2}$ peak of the non-hydrodynamic part of the front will eventually dominate the $t^{-3/4}$ peak due to quasiparticles. We do not see any sign of this in our numerics, but we do not have access to late enough times to address this asymptotic question. Whether the quasiparticles capture \emph{all} the operator weight, for some reason we do not yet understand, or whether there is instead a late-time crossover to a Gaussian front, is an interesting question for future work.

Table \ref{tab: table} summarizes the various scalings for the width and height of the operator weight for generic operators, in integrable, chaotic and non-interacting systems. We also note that our prediction for the operator front~\eqref{EqScaling} in interacting integrable systems also applies to the front of standard two-point correlation functions. Linear response correlation functions admit a hydrodynamic decomposition in terms of quasiparticles as in eq.~\eqref{Eq: toy}, so our argument carries over to such correlation functions. It will be interesting to check this prediction in future work. 

\section{Operator front  in chaotic systems}\label{sec: non-integrable}

\begin{figure}
\centering
\includegraphics[width=\linewidth,clip]{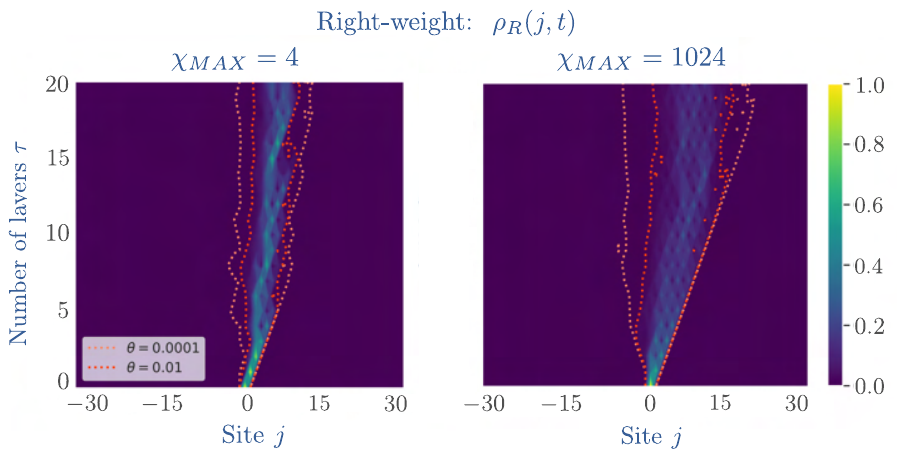}
 \caption{\textbf{Truncation errors on the right-weight for random circuit dynamics.} Here we show a single Haar random circuit realization. Left panel: $\chi_{\rm MAX}=4$. Right panel: $\chi_{\rm MAX}=1024$. The dashed lines are contour lines of the right-weight with threshold $\theta$. For small bond dimension, the operator front slows down, and stops at finite time. This is an artifact of truncation errors, that can be postponed to longer times by increasing the bond dimension. }
\label{Fig: heatmap_random}
\end{figure}

\begin{figure}
\centering
\includegraphics[width=\linewidth,clip]{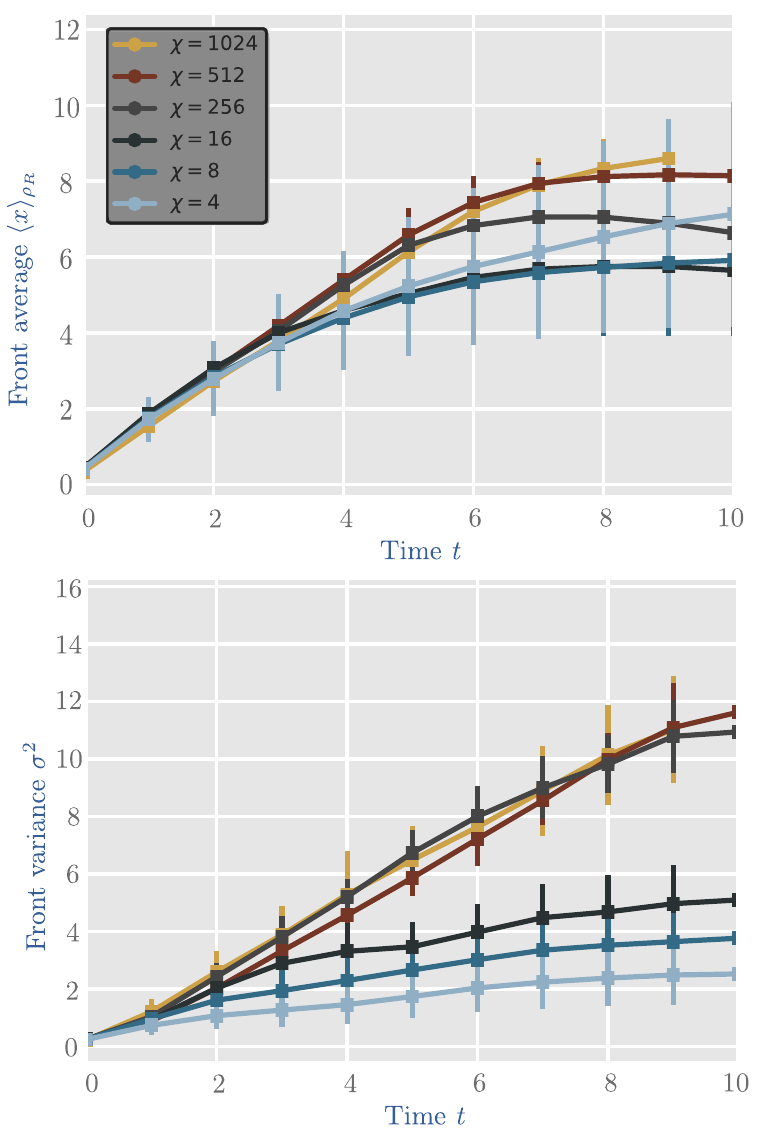}
 \caption{\textbf{Slow-down in operator spreading due to truncation errors.} Top panel: average front position vs time averaged over different circuit realizations and for various maximum bond dimensions $\chi_{\rm MAX}$. Bottom panel: same but for the front variance.}
\label{Fig: random_circuit_broadening}
\end{figure}

\begin{figure}
\centering
\includegraphics[width=\linewidth,clip]{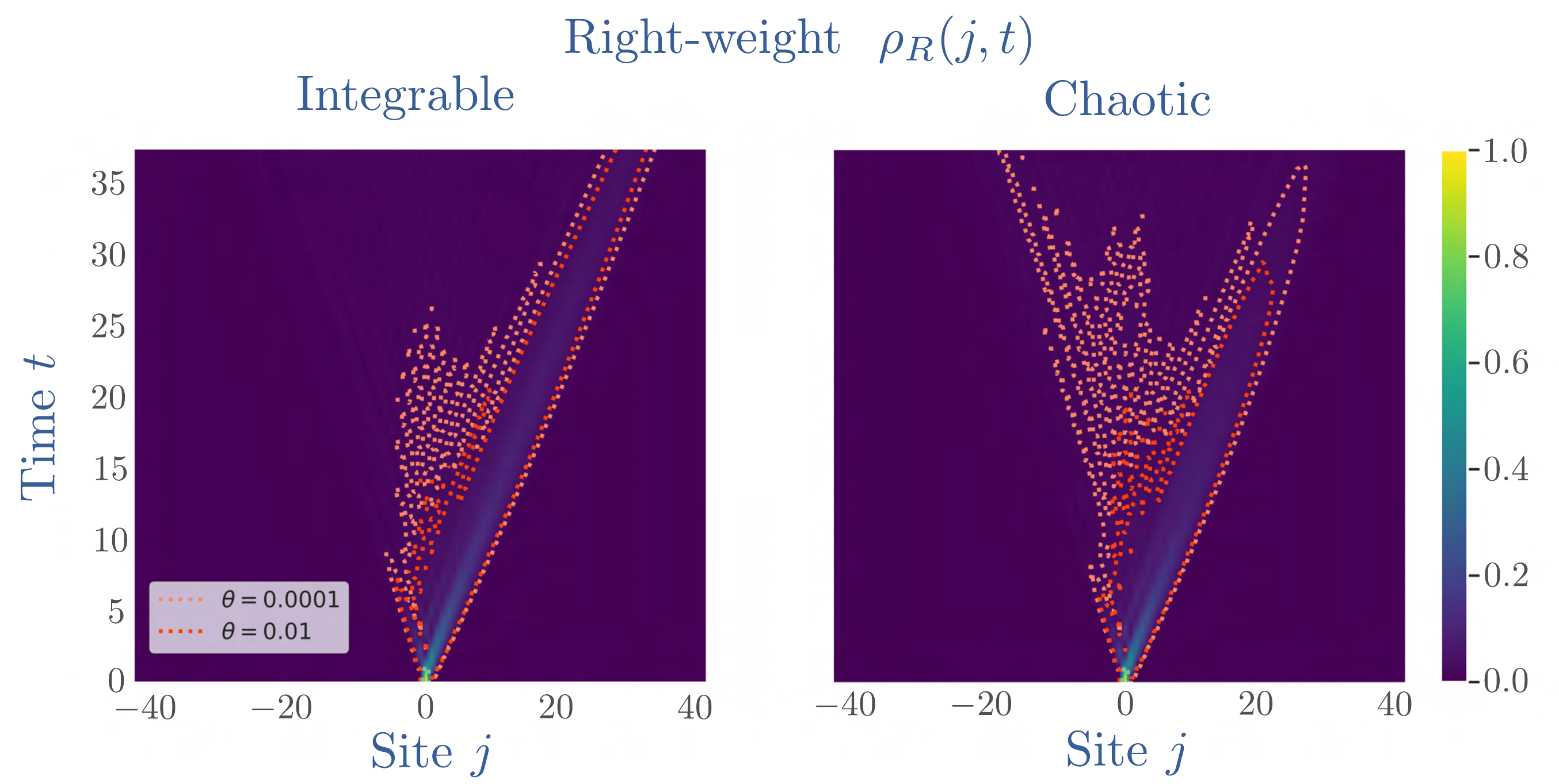}
 \caption{\textbf{Comparison of integrable vs. chaotic dynamics in operator spreading.} Left (Right): XXZ model at $\Delta=1/2$ and an homogeneous (staggered) magnetic field of $h_z=0.1$. Results at $\chi_{\rm MAX}=64$.}
\label{Fig: integrable_vs_non-integrable}
\end{figure}

We now briefly contrast our findings for interacting integrable systems to chaotic (non-integrable) chains. In chaotic systems, the operator front is expected to broaden diffusively~\cite{nahum2018operator,von2018operator} as in integrable systems  (albeit for very different reasons \cite{gopalakrishnan2018hydrodynamics}), but with a Gaussian scaling function. As we will show below, the effects of truncation errors using finite dimension MPOs are even more drastic for chaotic systems. In practice, this provides yet another way to distinguish integrability and chaos using finite bond dimension numerics, but this also makes accessing the true operator front properties of chaotic systems numerically very challenging.  

We first study random Haar quantum circuits where each two-site gate is independently drawn from the ensemble of Haar random matrices of size $q^2 \times q^2$, with $q$ the local Hilbert space dimension. Our results will focus on the case $q=2$ corresponding to spin$-1/2$ systems. The seminal works~\cite{nahum2018operator, von2018operator} analyzed this setup analytically and characterized operator spreading exactly. Our main motivation here is to study operator front broadening in this setup numerically, to illustrate the effects of truncation errors due to finite bond dimension. Our results indicate the following two features present in quantum chaotic models at finite bond dimensions: 1) artificial slow-down of operator spreading as shown in Figs. \ref{Fig: heatmap_random}-\ref{Fig: random_circuit_broadening} (see also Ref.~\onlinecite{PhysRevB.100.104303}); and 2) a front that broadens sub-diffusively and eventually stops broadening altogether, as shown in Fig. \ref{Fig: random_circuit_broadening} (see also~\cite{xu2020accessing} for similar results in the chaotic kicked Ising model). In Fig.~\ref{Fig: integrable_vs_non-integrable} we show how, even close to integrability, this slow-down in operator spreading becomes patent when studying the XXZ model for $\Delta=1/2$ and a staggered magnetic field along the $z-$ direction of $h_z=0.1$. Taking instead an homogeneous magnetic field of the same strength, in which case the system remains integrable, the front spreads ballistically at all times following the trace of the fastest quasiparticles in the system.

Those findings are consistent across all non-integrable models we have considered. We have also studied the chaotic Ising chain Hamiltonian given by:
\begin{equation}
H=J\sum_j \sigma_j^z\sigma_{j+1}^z+h_z\sigma_j^z+h_x\sigma_j^x.
\end{equation}
For simplicity we set again $J=1$. To ensure we are far into the non-integrable regime, we set $h_x=(1+\sqrt{5})/4$ and $h_z=(5+\sqrt{5})/8$, as in Ref.~\onlinecite{KimHuse}. Our simulations for the computation of the right-weight in this case require a time step $dt\lesssim 0.01$. In contrast with the integrable case analyzed in the previous section, the present non-integrable model yields a front that evades our MPO simulations entirely: the entire light cone structure vanishes after the maximum bond dimension is reached, after which the front fails to spread at all. We note that this phenomenon is absent in the integrable case (see middle panel in Fig. \ref{Fig: right-weight heatmap}). 
In fact, the maximum bond dimension in both the TFI model as well as in the XXZ model is reached at around the same time in both cases. This hints at a possible connection already put forward in Ref.~\onlinecite{PhysRevE.75.015202} between operator entanglement growth and integrability.

\section{Conclusion} \label{sec: conclusion} 
We have analyzed operator spreading in generic quantum many-body systems by computing the right-weight numerically, using matrix product operators. Contrary to earlier expectations, we have shown that correctly capturing the operator front and its broadening requires large bond dimensions, and that truncation errors can lead to erroneous conclusions. In chaotic systems, we find that the operator front is quickly out of reach even using large bond dimensions of order ${\cal O}(1000)$. 
While the operator front broadens diffusively for both chaotic and integrable systems, we identified a key difference in the precise shape of the front. For all times accessible to MPO calculations, the operator front in integrable systems couples to a continuum of quasiparticles with velocities close to $v_B = {\rm max}_{\lambda} v_\lambda$. As a result, we argued on general grounds that hydrodynamic contributions to the front are non-Gaussian and have a height that decays anomalously as $t^{-3/4}$. These contributions will be accompanied for generic operators by additional non-hydrodynamic contributions decaying as $t^{-1/2}$, which would dominate at asymptotically late times. However, our numerical simulations detect the hydrodynamic contributions for operator evolutions of the local charge densities and the presence of the non-hydrodynamic contributions is still an open question. This provides a signature of integrability in operator spreading, albeit a rather subtle one, which may not persist to asymptotically late times. Most other features of the OTOC or right-weight appear to be qualitatively and quantitively similar in integrable and chaotic systems. 

It would be interesting to identify more differences in the future. A promising candidate is the value of the saturation of the OTOC behind the front, which is universal in chaotic systems, but is likely different in integrable systems. Another interesting direction is to understand further operator spreading at the isotropic point $\Delta=1$ in the XXZ model (Heisenberg chain). Indeed, our numerical results for front broadening are consistent with diffusive broadening for all the integrable systems we have considered, except at the Heisenberg chain---see Appendix, where we observe a \emph{subdiffusive} front broadening. While the Heisenberg chain is known to exhibit anomalous spin transport~\cite{ljubotina2017spin, PhysRevLett.121.230602, PhysRevLett.122.127202}, there is reason to expect this should have any consequence on the operator front. In fact, the fastest quasiparticle in the XXX spin chain has a finite diffusion constant~\cite{privateCommJacopo}, suggesting diffusive broadening. We leave the resolution of this mystery to future work.


\emph{Acknowledgments}.---The authors thank Jacopo De Nardis for stimulating discussions, in particular about front broadening in the Heisenberg spin chain. We acknowledge support from NSF Grant No. DMR-1653271 (S.G.),  the Air Force Office of Scientific Research under Grant No.~FA9550-21-1-0123 (R.V), and the Alfred P. Sloan Foundation through a Sloan Research Fellowship (R.V.).


\appendix 

\begin{figure*}[t!]
\centering
\includegraphics[width=\linewidth,clip]{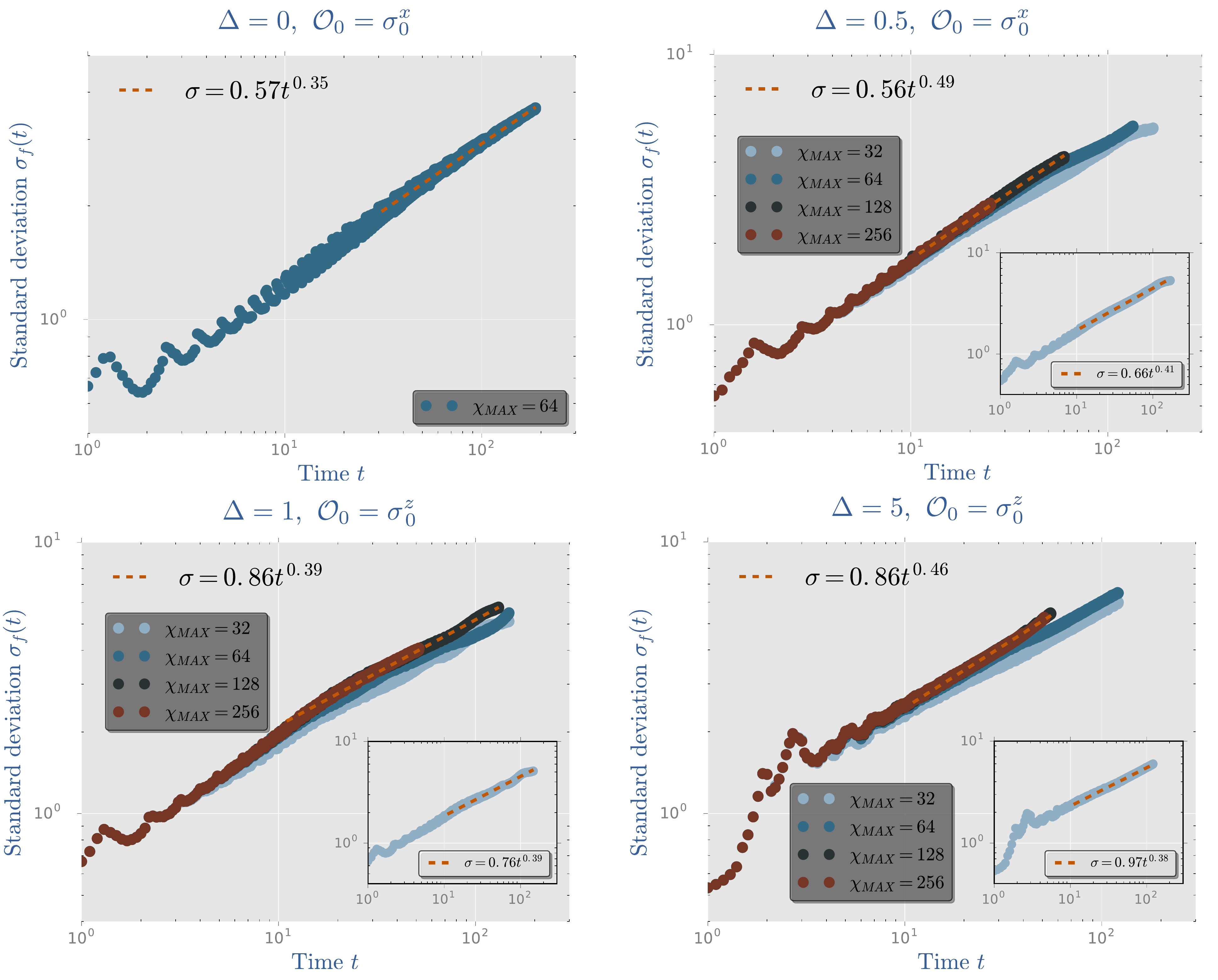}
\caption{\textbf{Operator front broadening in XXZ}. Operator front broadening in the XXZ model in the four possible regimes showing that large bond dimensions are required for convergence onto diffusive front broadening behavior (except right at $\Delta=1$): \textit{(i)} free fermions ($\Delta=0$); \textit{(ii)} gapless regime ($|\Delta|<1$); \textit{(iii)} isotropic point ($\Delta=1$); and gapped regime ($|\Delta|>1$).  Insets: results for $\chi_{\rm MAX}=32$.}
\label{Fig: spreading_various_Delta}
\end{figure*}

\section{Additional numerical results for the front broadening in integrable systems}

\begin{figure*}[t!]
\centering
\includegraphics[scale=0.4]{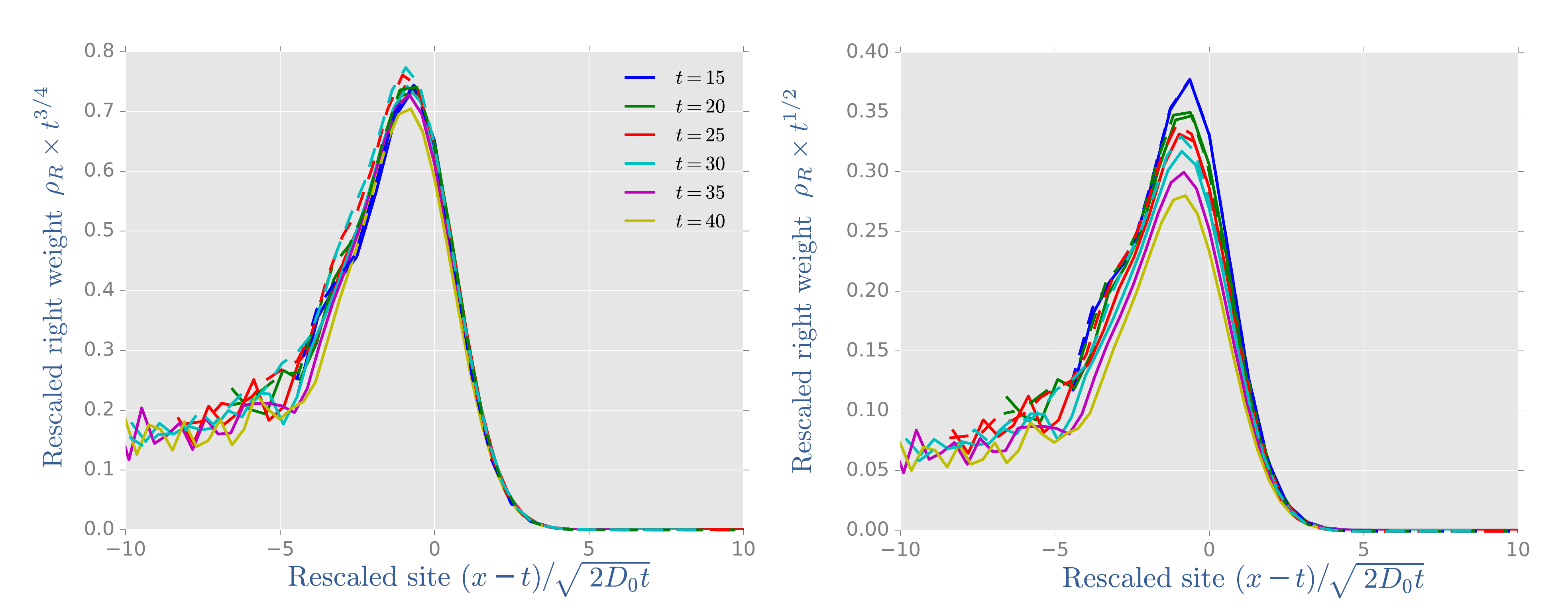}
\caption{\textbf{Benchmark of anomalous exponent for the right-weight in the integrable XXZ model for $\Delta=1/2$ against increasing bond dimension}.  Left figure: Scaling collapse of the front of the right-weight for the conserved operator $S_z$ for $\chi_{\rm MAX}=128$ ($\chi_{\rm MAX}=256$) solid lines (dashed lines) in agreement with Eq. (\ref{EqScaling}). Right figure: same data as in the left figure but multiplying the $y$-axis by $t^{1/2}$,  indicating that the front fails to collapse to a gaussian.}
\label{Fig: benchmark_anomalous}
\end{figure*}

In this section we provide further evidence of the subdiffusive spreading of the operator front generically in the integrable XXZ spin chain. In Fig. \ref{Fig: spreading_various_Delta}, we show how at finite bond dimension the right-weight spreads instead subdiffusively, approaching diffusive front broadening only at sufficiently large $\chi_{\rm MAX}$. While we numerically observe such a trend for almost all integrable models we have considered (see also Fig. \ref{Fig: benchmark_anomalous}), right at $\Delta=1$, our results seem to indicate instead a saturation to an anomalous exponent indicating subdiffusive broadening. We do not understand the reason for this subdiffusive broadening at the moment. From GHD calculations, the fastest quasiparticle in the isotropic Heisenberg spin chain is known to have a finite diffusion constant~\cite{privateCommJacopo}. It would be interesting to understand the reason for this apparent subdiffusive front broadening, as well as possible relations to operator entanglement growth~\cite{alba2020diffusion} in future works. 

\bibliography{refs_OTOC}

\begin{thebibliography}{82}%
\makeatletter
\providecommand \@ifxundefined [1]{%
 \@ifx{#1\undefined}
}%
\providecommand \@ifnum [1]{%
 \ifnum #1\expandafter \@firstoftwo
 \else \expandafter \@secondoftwo
 \fi
}%
\providecommand \@ifx [1]{%
 \ifx #1\expandafter \@firstoftwo
 \else \expandafter \@secondoftwo
 \fi
}%
\providecommand \natexlab [1]{#1}%
\providecommand \enquote  [1]{``#1''}%
\providecommand \bibnamefont  [1]{#1}%
\providecommand \bibfnamefont [1]{#1}%
\providecommand \citenamefont [1]{#1}%
\providecommand \href@noop [0]{\@secondoftwo}%
\providecommand \href [0]{\begingroup \@sanitize@url \@href}%
\providecommand \@href[1]{\@@startlink{#1}\@@href}%
\providecommand \@@href[1]{\endgroup#1\@@endlink}%
\providecommand \@sanitize@url [0]{\catcode `\\12\catcode `\$12\catcode
  `\&12\catcode `\#12\catcode `\^12\catcode `\_12\catcode `\%12\relax}%
\providecommand \@@startlink[1]{}%
\providecommand \@@endlink[0]{}%
\providecommand \url  [0]{\begingroup\@sanitize@url \@url }%
\providecommand \@url [1]{\endgroup\@href {#1}{\urlprefix }}%
\providecommand \urlprefix  [0]{URL }%
\providecommand \Eprint [0]{\href }%
\providecommand \doibase [0]{http://dx.doi.org/}%
\providecommand \selectlanguage [0]{\@gobble}%
\providecommand \bibinfo  [0]{\@secondoftwo}%
\providecommand \bibfield  [0]{\@secondoftwo}%
\providecommand \translation [1]{[#1]}%
\providecommand \BibitemOpen [0]{}%
\providecommand \bibitemStop [0]{}%
\providecommand \bibitemNoStop [0]{.\EOS\space}%
\providecommand \EOS [0]{\spacefactor3000\relax}%
\providecommand \BibitemShut  [1]{\csname bibitem#1\endcsname}%
\let\auto@bib@innerbib\@empty
\bibitem [{\citenamefont {Deutsch}(1991)}]{PhysRevA.43.2046}%
  \BibitemOpen
  \bibfield  {author} {\bibinfo {author} {\bibfnamefont {J.~M.}\ \bibnamefont
  {Deutsch}},\ }\href {\doibase 10.1103/PhysRevA.43.2046} {\bibfield  {journal}
  {\bibinfo  {journal} {Phys. Rev. A}\ }\textbf {\bibinfo {volume} {43}},\
  \bibinfo {pages} {2046} (\bibinfo {year} {1991})}\BibitemShut {NoStop}%
\bibitem [{\citenamefont {Srednicki}(1994)}]{PhysRevE.50.888}%
  \BibitemOpen
  \bibfield  {author} {\bibinfo {author} {\bibfnamefont {M.}~\bibnamefont
  {Srednicki}},\ }\href {\doibase 10.1103/PhysRevE.50.888} {\bibfield
  {journal} {\bibinfo  {journal} {Phys. Rev. E}\ }\textbf {\bibinfo {volume}
  {50}},\ \bibinfo {pages} {888} (\bibinfo {year} {1994})}\BibitemShut
  {NoStop}%
\bibitem [{\citenamefont {Goldstein}\ \emph {et~al.}(2010)\citenamefont
  {Goldstein}, \citenamefont {Lebowitz}, \citenamefont {Tumulka},\ and\
  \citenamefont {Zanghi}}]{goldstein2010long}%
  \BibitemOpen
  \bibfield  {author} {\bibinfo {author} {\bibfnamefont {S.}~\bibnamefont
  {Goldstein}}, \bibinfo {author} {\bibfnamefont {J.~L.}\ \bibnamefont
  {Lebowitz}}, \bibinfo {author} {\bibfnamefont {R.}~\bibnamefont {Tumulka}}, \
  and\ \bibinfo {author} {\bibfnamefont {N.}~\bibnamefont {Zanghi}},\
  }\href@noop {} {\bibfield  {journal} {\bibinfo  {journal} {The European
  Physical Journal H}\ }\textbf {\bibinfo {volume} {35}},\ \bibinfo {pages}
  {173} (\bibinfo {year} {2010})}\BibitemShut {NoStop}%
\bibitem [{\citenamefont {Hayden}\ and\ \citenamefont
  {Preskill}(2007)}]{hayden2007black}%
  \BibitemOpen
  \bibfield  {author} {\bibinfo {author} {\bibfnamefont {P.}~\bibnamefont
  {Hayden}}\ and\ \bibinfo {author} {\bibfnamefont {J.}~\bibnamefont
  {Preskill}},\ }\href@noop {} {\bibfield  {journal} {\bibinfo  {journal}
  {Journal of high energy physics}\ }\textbf {\bibinfo {volume} {2007}},\
  \bibinfo {pages} {120} (\bibinfo {year} {2007})}\BibitemShut {NoStop}%
\bibitem [{\citenamefont {Shenker}\ and\ \citenamefont
  {Stanford}(2014)}]{shenker2014black}%
  \BibitemOpen
  \bibfield  {author} {\bibinfo {author} {\bibfnamefont {S.~H.}\ \bibnamefont
  {Shenker}}\ and\ \bibinfo {author} {\bibfnamefont {D.}~\bibnamefont
  {Stanford}},\ }\href@noop {} {\bibfield  {journal} {\bibinfo  {journal}
  {Journal of High Energy Physics}\ }\textbf {\bibinfo {volume} {2014}},\
  \bibinfo {pages} {67} (\bibinfo {year} {2014})}\BibitemShut {NoStop}%
\bibitem [{\citenamefont {Trotzky}\ \emph {et~al.}(2012)\citenamefont
  {Trotzky}, \citenamefont {Chen}, \citenamefont {Flesch}, \citenamefont
  {McCulloch}, \citenamefont {Schollw{\"o}ck}, \citenamefont {Eisert},\ and\
  \citenamefont {Bloch}}]{trotzky2012probing}%
  \BibitemOpen
  \bibfield  {author} {\bibinfo {author} {\bibfnamefont {S.}~\bibnamefont
  {Trotzky}}, \bibinfo {author} {\bibfnamefont {Y.-A.}\ \bibnamefont {Chen}},
  \bibinfo {author} {\bibfnamefont {A.}~\bibnamefont {Flesch}}, \bibinfo
  {author} {\bibfnamefont {I.~P.}\ \bibnamefont {McCulloch}}, \bibinfo {author}
  {\bibfnamefont {U.}~\bibnamefont {Schollw{\"o}ck}}, \bibinfo {author}
  {\bibfnamefont {J.}~\bibnamefont {Eisert}}, \ and\ \bibinfo {author}
  {\bibfnamefont {I.}~\bibnamefont {Bloch}},\ }\href@noop {} {\bibfield
  {journal} {\bibinfo  {journal} {Nature physics}\ }\textbf {\bibinfo {volume}
  {8}},\ \bibinfo {pages} {325} (\bibinfo {year} {2012})}\BibitemShut {NoStop}%
\bibitem [{\citenamefont {Gring}\ \emph {et~al.}(2012)\citenamefont {Gring},
  \citenamefont {Kuhnert}, \citenamefont {Langen}, \citenamefont {Kitagawa},
  \citenamefont {Rauer}, \citenamefont {Schreitl}, \citenamefont {Mazets},
  \citenamefont {Smith}, \citenamefont {Demler},\ and\ \citenamefont
  {Schmiedmayer}}]{gring2012relaxation}%
  \BibitemOpen
  \bibfield  {author} {\bibinfo {author} {\bibfnamefont {M.}~\bibnamefont
  {Gring}}, \bibinfo {author} {\bibfnamefont {M.}~\bibnamefont {Kuhnert}},
  \bibinfo {author} {\bibfnamefont {T.}~\bibnamefont {Langen}}, \bibinfo
  {author} {\bibfnamefont {T.}~\bibnamefont {Kitagawa}}, \bibinfo {author}
  {\bibfnamefont {B.}~\bibnamefont {Rauer}}, \bibinfo {author} {\bibfnamefont
  {M.}~\bibnamefont {Schreitl}}, \bibinfo {author} {\bibfnamefont
  {I.}~\bibnamefont {Mazets}}, \bibinfo {author} {\bibfnamefont {D.~A.}\
  \bibnamefont {Smith}}, \bibinfo {author} {\bibfnamefont {E.}~\bibnamefont
  {Demler}}, \ and\ \bibinfo {author} {\bibfnamefont {J.}~\bibnamefont
  {Schmiedmayer}},\ }\href@noop {} {\bibfield  {journal} {\bibinfo  {journal}
  {Science}\ }\textbf {\bibinfo {volume} {337}},\ \bibinfo {pages} {1318}
  (\bibinfo {year} {2012})}\BibitemShut {NoStop}%
\bibitem [{\citenamefont {Schneider}\ \emph {et~al.}(2012)\citenamefont
  {Schneider}, \citenamefont {Hackerm{\"u}ller}, \citenamefont {Ronzheimer},
  \citenamefont {Will}, \citenamefont {Braun}, \citenamefont {Best},
  \citenamefont {Bloch}, \citenamefont {Demler}, \citenamefont {Mandt},
  \citenamefont {Rasch} \emph {et~al.}}]{schneider2012fermionic}%
  \BibitemOpen
  \bibfield  {author} {\bibinfo {author} {\bibfnamefont {U.}~\bibnamefont
  {Schneider}}, \bibinfo {author} {\bibfnamefont {L.}~\bibnamefont
  {Hackerm{\"u}ller}}, \bibinfo {author} {\bibfnamefont {J.~P.}\ \bibnamefont
  {Ronzheimer}}, \bibinfo {author} {\bibfnamefont {S.}~\bibnamefont {Will}},
  \bibinfo {author} {\bibfnamefont {S.}~\bibnamefont {Braun}}, \bibinfo
  {author} {\bibfnamefont {T.}~\bibnamefont {Best}}, \bibinfo {author}
  {\bibfnamefont {I.}~\bibnamefont {Bloch}}, \bibinfo {author} {\bibfnamefont
  {E.}~\bibnamefont {Demler}}, \bibinfo {author} {\bibfnamefont
  {S.}~\bibnamefont {Mandt}}, \bibinfo {author} {\bibfnamefont
  {D.}~\bibnamefont {Rasch}},  \emph {et~al.},\ }\href@noop {} {\bibfield
  {journal} {\bibinfo  {journal} {Nature Physics}\ }\textbf {\bibinfo {volume}
  {8}},\ \bibinfo {pages} {213} (\bibinfo {year} {2012})}\BibitemShut {NoStop}%
\bibitem [{\citenamefont {Cheneau}\ \emph {et~al.}(2012)\citenamefont
  {Cheneau}, \citenamefont {Barmettler}, \citenamefont {Poletti}, \citenamefont
  {Endres}, \citenamefont {Schau{\ss}}, \citenamefont {Fukuhara}, \citenamefont
  {Gross}, \citenamefont {Bloch}, \citenamefont {Kollath},\ and\ \citenamefont
  {Kuhr}}]{cheneau2012light}%
  \BibitemOpen
  \bibfield  {author} {\bibinfo {author} {\bibfnamefont {M.}~\bibnamefont
  {Cheneau}}, \bibinfo {author} {\bibfnamefont {P.}~\bibnamefont {Barmettler}},
  \bibinfo {author} {\bibfnamefont {D.}~\bibnamefont {Poletti}}, \bibinfo
  {author} {\bibfnamefont {M.}~\bibnamefont {Endres}}, \bibinfo {author}
  {\bibfnamefont {P.}~\bibnamefont {Schau{\ss}}}, \bibinfo {author}
  {\bibfnamefont {T.}~\bibnamefont {Fukuhara}}, \bibinfo {author}
  {\bibfnamefont {C.}~\bibnamefont {Gross}}, \bibinfo {author} {\bibfnamefont
  {I.}~\bibnamefont {Bloch}}, \bibinfo {author} {\bibfnamefont
  {C.}~\bibnamefont {Kollath}}, \ and\ \bibinfo {author} {\bibfnamefont
  {S.}~\bibnamefont {Kuhr}},\ }\href@noop {} {\bibfield  {journal} {\bibinfo
  {journal} {Nature}\ }\textbf {\bibinfo {volume} {481}},\ \bibinfo {pages}
  {484} (\bibinfo {year} {2012})}\BibitemShut {NoStop}%
\bibitem [{\citenamefont {Islam}\ \emph {et~al.}(2015)\citenamefont {Islam},
  \citenamefont {Ma}, \citenamefont {Preiss}, \citenamefont {Eric~Tai},
  \citenamefont {Lukin}, \citenamefont {Rispoli},\ and\ \citenamefont
  {Greiner}}]{Islam:2015aa}%
  \BibitemOpen
  \bibfield  {author} {\bibinfo {author} {\bibfnamefont {R.}~\bibnamefont
  {Islam}}, \bibinfo {author} {\bibfnamefont {R.}~\bibnamefont {Ma}}, \bibinfo
  {author} {\bibfnamefont {P.~M.}\ \bibnamefont {Preiss}}, \bibinfo {author}
  {\bibfnamefont {M.}~\bibnamefont {Eric~Tai}}, \bibinfo {author}
  {\bibfnamefont {A.}~\bibnamefont {Lukin}}, \bibinfo {author} {\bibfnamefont
  {M.}~\bibnamefont {Rispoli}}, \ and\ \bibinfo {author} {\bibfnamefont
  {M.}~\bibnamefont {Greiner}},\ }\href {\doibase 10.1038/nature15750}
  {\bibfield  {journal} {\bibinfo  {journal} {Nature}\ }\textbf {\bibinfo
  {volume} {528}},\ \bibinfo {pages} {77} (\bibinfo {year} {2015})}\BibitemShut
  {NoStop}%
\bibitem [{\citenamefont {Lukin}\ \emph {et~al.}(2019)\citenamefont {Lukin},
  \citenamefont {Rispoli}, \citenamefont {Schittko}, \citenamefont {Tai},
  \citenamefont {Kaufman}, \citenamefont {Choi}, \citenamefont {Khemani},
  \citenamefont {L{\'e}onard},\ and\ \citenamefont {Greiner}}]{Lukin256}%
  \BibitemOpen
  \bibfield  {author} {\bibinfo {author} {\bibfnamefont {A.}~\bibnamefont
  {Lukin}}, \bibinfo {author} {\bibfnamefont {M.}~\bibnamefont {Rispoli}},
  \bibinfo {author} {\bibfnamefont {R.}~\bibnamefont {Schittko}}, \bibinfo
  {author} {\bibfnamefont {M.~E.}\ \bibnamefont {Tai}}, \bibinfo {author}
  {\bibfnamefont {A.~M.}\ \bibnamefont {Kaufman}}, \bibinfo {author}
  {\bibfnamefont {S.}~\bibnamefont {Choi}}, \bibinfo {author} {\bibfnamefont
  {V.}~\bibnamefont {Khemani}}, \bibinfo {author} {\bibfnamefont
  {J.}~\bibnamefont {L{\'e}onard}}, \ and\ \bibinfo {author} {\bibfnamefont
  {M.}~\bibnamefont {Greiner}},\ }\href {\doibase 10.1126/science.aau0818}
  {\bibfield  {journal} {\bibinfo  {journal} {Science}\ }\textbf {\bibinfo
  {volume} {364}},\ \bibinfo {pages} {256} (\bibinfo {year} {2019})},\ \Eprint
  {http://arxiv.org/abs/https://science.sciencemag.org/content/364/6437/256.full.pdf}
  {https://science.sciencemag.org/content/364/6437/256.full.pdf} \BibitemShut
  {NoStop}%
\bibitem [{\citenamefont {{Chiaro}}\ \emph {et~al.}(2019)\citenamefont
  {{Chiaro}}, \citenamefont {{Neill}}, \citenamefont {{Bohrdt}}, \citenamefont
  {{Filippone}}, \citenamefont {{Arute}}, \citenamefont {{Arya}}, \citenamefont
  {{Babbush}}, \citenamefont {{Bacon}}, \citenamefont {{Bardin}}, \citenamefont
  {{Barends}}, \citenamefont {{Boixo}}, \citenamefont {{Buell}}, \citenamefont
  {{Burkett}}, \citenamefont {{Chen}}, \citenamefont {{Chen}}, \citenamefont
  {{Collins}}, \citenamefont {{Dunsworth}}, \citenamefont {{Farhi}},
  \citenamefont {{Fowler}}, \citenamefont {{Foxen}}, \citenamefont {{Gidney}},
  \citenamefont {{Giustina}}, \citenamefont {{Harrigan}}, \citenamefont
  {{Huang}}, \citenamefont {{Isakov}}, \citenamefont {{Jeffrey}}, \citenamefont
  {{Jiang}}, \citenamefont {{Kafri}}, \citenamefont {{Kechedzhi}},
  \citenamefont {{Kelly}}, \citenamefont {{Klimov}}, \citenamefont
  {{Korotkov}}, \citenamefont {{Kostritsa}}, \citenamefont {{Landhuis}},
  \citenamefont {{Lucero}}, \citenamefont {{McClean}}, \citenamefont {{Mi}},
  \citenamefont {{Megrant}}, \citenamefont {{Mohseni}}, \citenamefont
  {{Mutus}}, \citenamefont {{McEwen}}, \citenamefont {{Naaman}}, \citenamefont
  {{Neeley}}, \citenamefont {{Niu}}, \citenamefont {{Petukhov}}, \citenamefont
  {{Quintana}}, \citenamefont {{Rubin}}, \citenamefont {{Sank}}, \citenamefont
  {{Satzinger}}, \citenamefont {{Vainsencher}}, \citenamefont {{White}},
  \citenamefont {{Yao}}, \citenamefont {{Yeh}}, \citenamefont {{Zalcman}},
  \citenamefont {{Smelyanskiy}}, \citenamefont {{Neven}}, \citenamefont
  {{Gopalakrishnan}}, \citenamefont {{Abanin}}, \citenamefont {{Knap}},
  \citenamefont {{Martinis}},\ and\ \citenamefont
  {{Roushan}}}]{2019arXiv191006024C}%
  \BibitemOpen
  \bibfield  {author} {\bibinfo {author} {\bibfnamefont {B.}~\bibnamefont
  {{Chiaro}}}, \bibinfo {author} {\bibfnamefont {C.}~\bibnamefont {{Neill}}},
  \bibinfo {author} {\bibfnamefont {A.}~\bibnamefont {{Bohrdt}}}, \bibinfo
  {author} {\bibfnamefont {M.}~\bibnamefont {{Filippone}}}, \bibinfo {author}
  {\bibfnamefont {F.}~\bibnamefont {{Arute}}}, \bibinfo {author} {\bibfnamefont
  {K.}~\bibnamefont {{Arya}}}, \bibinfo {author} {\bibfnamefont
  {R.}~\bibnamefont {{Babbush}}}, \bibinfo {author} {\bibfnamefont
  {D.}~\bibnamefont {{Bacon}}}, \bibinfo {author} {\bibfnamefont
  {J.}~\bibnamefont {{Bardin}}}, \bibinfo {author} {\bibfnamefont
  {R.}~\bibnamefont {{Barends}}}, \bibinfo {author} {\bibfnamefont
  {S.}~\bibnamefont {{Boixo}}}, \bibinfo {author} {\bibfnamefont
  {D.}~\bibnamefont {{Buell}}}, \bibinfo {author} {\bibfnamefont
  {B.}~\bibnamefont {{Burkett}}}, \bibinfo {author} {\bibfnamefont
  {Y.}~\bibnamefont {{Chen}}}, \bibinfo {author} {\bibfnamefont
  {Z.}~\bibnamefont {{Chen}}}, \bibinfo {author} {\bibfnamefont
  {R.}~\bibnamefont {{Collins}}}, \bibinfo {author} {\bibfnamefont
  {A.}~\bibnamefont {{Dunsworth}}}, \bibinfo {author} {\bibfnamefont
  {E.}~\bibnamefont {{Farhi}}}, \bibinfo {author} {\bibfnamefont
  {A.}~\bibnamefont {{Fowler}}}, \bibinfo {author} {\bibfnamefont
  {B.}~\bibnamefont {{Foxen}}}, \bibinfo {author} {\bibfnamefont
  {C.}~\bibnamefont {{Gidney}}}, \bibinfo {author} {\bibfnamefont
  {M.}~\bibnamefont {{Giustina}}}, \bibinfo {author} {\bibfnamefont
  {M.}~\bibnamefont {{Harrigan}}}, \bibinfo {author} {\bibfnamefont
  {T.}~\bibnamefont {{Huang}}}, \bibinfo {author} {\bibfnamefont
  {S.}~\bibnamefont {{Isakov}}}, \bibinfo {author} {\bibfnamefont
  {E.}~\bibnamefont {{Jeffrey}}}, \bibinfo {author} {\bibfnamefont
  {Z.}~\bibnamefont {{Jiang}}}, \bibinfo {author} {\bibfnamefont
  {D.}~\bibnamefont {{Kafri}}}, \bibinfo {author} {\bibfnamefont
  {K.}~\bibnamefont {{Kechedzhi}}}, \bibinfo {author} {\bibfnamefont
  {J.}~\bibnamefont {{Kelly}}}, \bibinfo {author} {\bibfnamefont
  {P.}~\bibnamefont {{Klimov}}}, \bibinfo {author} {\bibfnamefont
  {A.}~\bibnamefont {{Korotkov}}}, \bibinfo {author} {\bibfnamefont
  {F.}~\bibnamefont {{Kostritsa}}}, \bibinfo {author} {\bibfnamefont
  {D.}~\bibnamefont {{Landhuis}}}, \bibinfo {author} {\bibfnamefont
  {E.}~\bibnamefont {{Lucero}}}, \bibinfo {author} {\bibfnamefont
  {J.}~\bibnamefont {{McClean}}}, \bibinfo {author} {\bibfnamefont
  {X.}~\bibnamefont {{Mi}}}, \bibinfo {author} {\bibfnamefont {A.}~\bibnamefont
  {{Megrant}}}, \bibinfo {author} {\bibfnamefont {M.}~\bibnamefont
  {{Mohseni}}}, \bibinfo {author} {\bibfnamefont {J.}~\bibnamefont {{Mutus}}},
  \bibinfo {author} {\bibfnamefont {M.}~\bibnamefont {{McEwen}}}, \bibinfo
  {author} {\bibfnamefont {O.}~\bibnamefont {{Naaman}}}, \bibinfo {author}
  {\bibfnamefont {M.}~\bibnamefont {{Neeley}}}, \bibinfo {author}
  {\bibfnamefont {M.}~\bibnamefont {{Niu}}}, \bibinfo {author} {\bibfnamefont
  {A.}~\bibnamefont {{Petukhov}}}, \bibinfo {author} {\bibfnamefont
  {C.}~\bibnamefont {{Quintana}}}, \bibinfo {author} {\bibfnamefont
  {N.}~\bibnamefont {{Rubin}}}, \bibinfo {author} {\bibfnamefont
  {D.}~\bibnamefont {{Sank}}}, \bibinfo {author} {\bibfnamefont
  {K.}~\bibnamefont {{Satzinger}}}, \bibinfo {author} {\bibfnamefont
  {A.}~\bibnamefont {{Vainsencher}}}, \bibinfo {author} {\bibfnamefont
  {T.}~\bibnamefont {{White}}}, \bibinfo {author} {\bibfnamefont
  {Z.}~\bibnamefont {{Yao}}}, \bibinfo {author} {\bibfnamefont
  {P.}~\bibnamefont {{Yeh}}}, \bibinfo {author} {\bibfnamefont
  {A.}~\bibnamefont {{Zalcman}}}, \bibinfo {author} {\bibfnamefont
  {V.}~\bibnamefont {{Smelyanskiy}}}, \bibinfo {author} {\bibfnamefont
  {H.}~\bibnamefont {{Neven}}}, \bibinfo {author} {\bibfnamefont
  {S.}~\bibnamefont {{Gopalakrishnan}}}, \bibinfo {author} {\bibfnamefont
  {D.}~\bibnamefont {{Abanin}}}, \bibinfo {author} {\bibfnamefont
  {M.}~\bibnamefont {{Knap}}}, \bibinfo {author} {\bibfnamefont
  {J.}~\bibnamefont {{Martinis}}}, \ and\ \bibinfo {author} {\bibfnamefont
  {P.}~\bibnamefont {{Roushan}}},\ }\href@noop {} {\bibfield  {journal}
  {\bibinfo  {journal} {arXiv e-prints}\ ,\ \bibinfo {eid} {arXiv:1910.06024}}
  (\bibinfo {year} {2019})},\ \Eprint {http://arxiv.org/abs/1910.06024}
  {arXiv:1910.06024 [cond-mat.dis-nn]} \BibitemShut {NoStop}%
\bibitem [{\citenamefont {Sekino}\ and\ \citenamefont
  {Susskind}(2008)}]{sekino2008fast}%
  \BibitemOpen
  \bibfield  {author} {\bibinfo {author} {\bibfnamefont {Y.}~\bibnamefont
  {Sekino}}\ and\ \bibinfo {author} {\bibfnamefont {L.}~\bibnamefont
  {Susskind}},\ }\href@noop {} {\bibfield  {journal} {\bibinfo  {journal}
  {Journal of High Energy Physics}\ }\textbf {\bibinfo {volume} {2008}},\
  \bibinfo {pages} {065} (\bibinfo {year} {2008})}\BibitemShut {NoStop}%
\bibitem [{\citenamefont {Hosur}\ \emph {et~al.}(2016)\citenamefont {Hosur},
  \citenamefont {Qi}, \citenamefont {Roberts},\ and\ \citenamefont
  {Yoshida}}]{hosur2016chaos}%
  \BibitemOpen
  \bibfield  {author} {\bibinfo {author} {\bibfnamefont {P.}~\bibnamefont
  {Hosur}}, \bibinfo {author} {\bibfnamefont {X.-L.}\ \bibnamefont {Qi}},
  \bibinfo {author} {\bibfnamefont {D.~A.}\ \bibnamefont {Roberts}}, \ and\
  \bibinfo {author} {\bibfnamefont {B.}~\bibnamefont {Yoshida}},\ }\href@noop
  {} {\bibfield  {journal} {\bibinfo  {journal} {Journal of High Energy
  Physics}\ }\textbf {\bibinfo {volume} {2016}},\ \bibinfo {pages} {4}
  (\bibinfo {year} {2016})}\BibitemShut {NoStop}%
\bibitem [{\citenamefont {Lashkari}\ \emph {et~al.}(2013)\citenamefont
  {Lashkari}, \citenamefont {Stanford}, \citenamefont {Hastings}, \citenamefont
  {Osborne},\ and\ \citenamefont {Hayden}}]{lashkari2013towards}%
  \BibitemOpen
  \bibfield  {author} {\bibinfo {author} {\bibfnamefont {N.}~\bibnamefont
  {Lashkari}}, \bibinfo {author} {\bibfnamefont {D.}~\bibnamefont {Stanford}},
  \bibinfo {author} {\bibfnamefont {M.}~\bibnamefont {Hastings}}, \bibinfo
  {author} {\bibfnamefont {T.}~\bibnamefont {Osborne}}, \ and\ \bibinfo
  {author} {\bibfnamefont {P.}~\bibnamefont {Hayden}},\ }\href@noop {}
  {\bibfield  {journal} {\bibinfo  {journal} {Journal of High Energy Physics}\
  }\textbf {\bibinfo {volume} {2013}},\ \bibinfo {pages} {22} (\bibinfo {year}
  {2013})}\BibitemShut {NoStop}%
\bibitem [{\citenamefont {Xu}\ and\ \citenamefont
  {Swingle}(2019)}]{PhysRevX.9.031048}%
  \BibitemOpen
  \bibfield  {author} {\bibinfo {author} {\bibfnamefont {S.}~\bibnamefont
  {Xu}}\ and\ \bibinfo {author} {\bibfnamefont {B.}~\bibnamefont {Swingle}},\
  }\href {\doibase 10.1103/PhysRevX.9.031048} {\bibfield  {journal} {\bibinfo
  {journal} {Phys. Rev. X}\ }\textbf {\bibinfo {volume} {9}},\ \bibinfo {pages}
  {031048} (\bibinfo {year} {2019})}\BibitemShut {NoStop}%
\bibitem [{\citenamefont {Parker}\ \emph {et~al.}(2019)\citenamefont {Parker},
  \citenamefont {Cao}, \citenamefont {Avdoshkin}, \citenamefont {Scaffidi},\
  and\ \citenamefont {Altman}}]{PhysRevX.9.041017}%
  \BibitemOpen
  \bibfield  {author} {\bibinfo {author} {\bibfnamefont {D.~E.}\ \bibnamefont
  {Parker}}, \bibinfo {author} {\bibfnamefont {X.}~\bibnamefont {Cao}},
  \bibinfo {author} {\bibfnamefont {A.}~\bibnamefont {Avdoshkin}}, \bibinfo
  {author} {\bibfnamefont {T.}~\bibnamefont {Scaffidi}}, \ and\ \bibinfo
  {author} {\bibfnamefont {E.}~\bibnamefont {Altman}},\ }\href {\doibase
  10.1103/PhysRevX.9.041017} {\bibfield  {journal} {\bibinfo  {journal} {Phys.
  Rev. X}\ }\textbf {\bibinfo {volume} {9}},\ \bibinfo {pages} {041017}
  (\bibinfo {year} {2019})}\BibitemShut {NoStop}%
\bibitem [{\citenamefont {Couch}\ \emph {et~al.}(2020)\citenamefont {Couch},
  \citenamefont {Eccles}, \citenamefont {Nguyen}, \citenamefont {Swingle},\
  and\ \citenamefont {Xu}}]{PhysRevB.102.045114}%
  \BibitemOpen
  \bibfield  {author} {\bibinfo {author} {\bibfnamefont {J.}~\bibnamefont
  {Couch}}, \bibinfo {author} {\bibfnamefont {S.}~\bibnamefont {Eccles}},
  \bibinfo {author} {\bibfnamefont {P.}~\bibnamefont {Nguyen}}, \bibinfo
  {author} {\bibfnamefont {B.}~\bibnamefont {Swingle}}, \ and\ \bibinfo
  {author} {\bibfnamefont {S.}~\bibnamefont {Xu}},\ }\href {\doibase
  10.1103/PhysRevB.102.045114} {\bibfield  {journal} {\bibinfo  {journal}
  {Phys. Rev. B}\ }\textbf {\bibinfo {volume} {102}},\ \bibinfo {pages}
  {045114} (\bibinfo {year} {2020})}\BibitemShut {NoStop}%
\bibitem [{\citenamefont {Lieb}\ and\ \citenamefont
  {Robinson}(1972)}]{cmp/1103858407}%
  \BibitemOpen
  \bibfield  {author} {\bibinfo {author} {\bibfnamefont {E.~H.}\ \bibnamefont
  {Lieb}}\ and\ \bibinfo {author} {\bibfnamefont {D.~W.}\ \bibnamefont
  {Robinson}},\ }\href {\doibase cmp/1103858407} {\bibfield  {journal}
  {\bibinfo  {journal} {Communications in Mathematical Physics}\ }\textbf
  {\bibinfo {volume} {28}},\ \bibinfo {pages} {251 } (\bibinfo {year}
  {1972})}\BibitemShut {NoStop}%
\bibitem [{\citenamefont {Kim}\ and\ \citenamefont {Huse}(2013)}]{KimHuse}%
  \BibitemOpen
  \bibfield  {author} {\bibinfo {author} {\bibfnamefont {H.}~\bibnamefont
  {Kim}}\ and\ \bibinfo {author} {\bibfnamefont {D.~A.}\ \bibnamefont {Huse}},\
  }\href {\doibase 10.1103/PhysRevLett.111.127205} {\bibfield  {journal}
  {\bibinfo  {journal} {Phys. Rev. Lett.}\ }\textbf {\bibinfo {volume} {111}},\
  \bibinfo {pages} {127205} (\bibinfo {year} {2013})}\BibitemShut {NoStop}%
\bibitem [{\citenamefont {{Bohrdt}}\ \emph {et~al.}(2017)\citenamefont
  {{Bohrdt}}, \citenamefont {{Mendl}}, \citenamefont {{Endres}},\ and\
  \citenamefont {{Knap}}}]{KnapScrambling}%
  \BibitemOpen
  \bibfield  {author} {\bibinfo {author} {\bibfnamefont {A.}~\bibnamefont
  {{Bohrdt}}}, \bibinfo {author} {\bibfnamefont {C.~B.}\ \bibnamefont
  {{Mendl}}}, \bibinfo {author} {\bibfnamefont {M.}~\bibnamefont {{Endres}}}, \
  and\ \bibinfo {author} {\bibfnamefont {M.}~\bibnamefont {{Knap}}},\ }\href
  {\doibase 10.1088/1367-2630/aa719b} {\bibfield  {journal} {\bibinfo
  {journal} {New Journal of Physics}\ }\textbf {\bibinfo {volume} {19}},\
  \bibinfo {eid} {063001} (\bibinfo {year} {2017})},\ \Eprint
  {http://arxiv.org/abs/1612.02434} {arXiv:1612.02434 [cond-mat.quant-gas]}
  \BibitemShut {NoStop}%
\bibitem [{\citenamefont {{Luitz}}\ and\ \citenamefont {{Bar
  Lev}}(2017)}]{LuitzScrambling}%
  \BibitemOpen
  \bibfield  {author} {\bibinfo {author} {\bibfnamefont {D.~J.}\ \bibnamefont
  {{Luitz}}}\ and\ \bibinfo {author} {\bibfnamefont {Y.}~\bibnamefont {{Bar
  Lev}}},\ }\href {\doibase 10.1103/PhysRevB.96.020406} {\bibfield  {journal}
  {\bibinfo  {journal} {\prb}\ }\textbf {\bibinfo {volume} {96}},\ \bibinfo
  {eid} {020406} (\bibinfo {year} {2017})},\ \Eprint
  {http://arxiv.org/abs/1702.03929} {arXiv:1702.03929 [cond-mat.dis-nn]}
  \BibitemShut {NoStop}%
\bibitem [{\citenamefont {Khemani}\ \emph
  {et~al.}(2018{\natexlab{a}})\citenamefont {Khemani}, \citenamefont
  {Vishwanath},\ and\ \citenamefont {Huse}}]{kvh}%
  \BibitemOpen
  \bibfield  {author} {\bibinfo {author} {\bibfnamefont {V.}~\bibnamefont
  {Khemani}}, \bibinfo {author} {\bibfnamefont {A.}~\bibnamefont {Vishwanath}},
  \ and\ \bibinfo {author} {\bibfnamefont {D.~A.}\ \bibnamefont {Huse}},\
  }\href {\doibase 10.1103/PhysRevX.8.031057} {\bibfield  {journal} {\bibinfo
  {journal} {Phys. Rev. X}\ }\textbf {\bibinfo {volume} {8}},\ \bibinfo {pages}
  {031057} (\bibinfo {year} {2018}{\natexlab{a}})}\BibitemShut {NoStop}%
\bibitem [{\citenamefont {Rakovszky}\ \emph {et~al.}(2018)\citenamefont
  {Rakovszky}, \citenamefont {Pollmann},\ and\ \citenamefont {von
  Keyserlingk}}]{PhysRevX.8.031058}%
  \BibitemOpen
  \bibfield  {author} {\bibinfo {author} {\bibfnamefont {T.}~\bibnamefont
  {Rakovszky}}, \bibinfo {author} {\bibfnamefont {F.}~\bibnamefont {Pollmann}},
  \ and\ \bibinfo {author} {\bibfnamefont {C.~W.}\ \bibnamefont {von
  Keyserlingk}},\ }\href {\doibase 10.1103/PhysRevX.8.031058} {\bibfield
  {journal} {\bibinfo  {journal} {Phys. Rev. X}\ }\textbf {\bibinfo {volume}
  {8}},\ \bibinfo {pages} {031058} (\bibinfo {year} {2018})}\BibitemShut
  {NoStop}%
\bibitem [{\citenamefont {Larkin}\ and\ \citenamefont
  {Ovchinnikov}(1969)}]{larkin1969quasiclassical}%
  \BibitemOpen
  \bibfield  {author} {\bibinfo {author} {\bibfnamefont {A.}~\bibnamefont
  {Larkin}}\ and\ \bibinfo {author} {\bibfnamefont {Y.~N.}\ \bibnamefont
  {Ovchinnikov}},\ }\href@noop {} {\bibfield  {journal} {\bibinfo  {journal}
  {Sov Phys JETP}\ }\textbf {\bibinfo {volume} {28}},\ \bibinfo {pages} {1200}
  (\bibinfo {year} {1969})}\BibitemShut {NoStop}%
\bibitem [{\citenamefont {Maldacena}\ \emph {et~al.}(2016)\citenamefont
  {Maldacena}, \citenamefont {Shenker},\ and\ \citenamefont
  {Stanford}}]{maldacena2016bound}%
  \BibitemOpen
  \bibfield  {author} {\bibinfo {author} {\bibfnamefont {J.}~\bibnamefont
  {Maldacena}}, \bibinfo {author} {\bibfnamefont {S.~H.}\ \bibnamefont
  {Shenker}}, \ and\ \bibinfo {author} {\bibfnamefont {D.}~\bibnamefont
  {Stanford}},\ }\href@noop {} {\bibfield  {journal} {\bibinfo  {journal}
  {Journal of High Energy Physics}\ }\textbf {\bibinfo {volume} {2016}},\
  \bibinfo {pages} {106} (\bibinfo {year} {2016})}\BibitemShut {NoStop}%
\bibitem [{\citenamefont {Fan}\ \emph {et~al.}(2017)\citenamefont {Fan},
  \citenamefont {Zhang}, \citenamefont {Shen},\ and\ \citenamefont
  {Zhai}}]{fan2017out}%
  \BibitemOpen
  \bibfield  {author} {\bibinfo {author} {\bibfnamefont {R.}~\bibnamefont
  {Fan}}, \bibinfo {author} {\bibfnamefont {P.}~\bibnamefont {Zhang}}, \bibinfo
  {author} {\bibfnamefont {H.}~\bibnamefont {Shen}}, \ and\ \bibinfo {author}
  {\bibfnamefont {H.}~\bibnamefont {Zhai}},\ }\href@noop {} {\bibfield
  {journal} {\bibinfo  {journal} {Science bulletin}\ }\textbf {\bibinfo
  {volume} {62}},\ \bibinfo {pages} {707} (\bibinfo {year} {2017})}\BibitemShut
  {NoStop}%
\bibitem [{\citenamefont {Swingle}\ and\ \citenamefont
  {Chowdhury}(2017)}]{PhysRevB.95.060201}%
  \BibitemOpen
  \bibfield  {author} {\bibinfo {author} {\bibfnamefont {B.}~\bibnamefont
  {Swingle}}\ and\ \bibinfo {author} {\bibfnamefont {D.}~\bibnamefont
  {Chowdhury}},\ }\href {\doibase 10.1103/PhysRevB.95.060201} {\bibfield
  {journal} {\bibinfo  {journal} {Phys. Rev. B}\ }\textbf {\bibinfo {volume}
  {95}},\ \bibinfo {pages} {060201} (\bibinfo {year} {2017})}\BibitemShut
  {NoStop}%
\bibitem [{\citenamefont {Chen}\ \emph {et~al.}(2017)\citenamefont {Chen},
  \citenamefont {Zhou}, \citenamefont {Huse},\ and\ \citenamefont
  {Fradkin}}]{chen2017out}%
  \BibitemOpen
  \bibfield  {author} {\bibinfo {author} {\bibfnamefont {X.}~\bibnamefont
  {Chen}}, \bibinfo {author} {\bibfnamefont {T.}~\bibnamefont {Zhou}}, \bibinfo
  {author} {\bibfnamefont {D.~A.}\ \bibnamefont {Huse}}, \ and\ \bibinfo
  {author} {\bibfnamefont {E.}~\bibnamefont {Fradkin}},\ }\href@noop {}
  {\bibfield  {journal} {\bibinfo  {journal} {Annalen der Physik}\ }\textbf
  {\bibinfo {volume} {529}},\ \bibinfo {pages} {1600332} (\bibinfo {year}
  {2017})}\BibitemShut {NoStop}%
\bibitem [{\citenamefont {Von~Keyserlingk}\ \emph {et~al.}(2018)\citenamefont
  {Von~Keyserlingk}, \citenamefont {Rakovszky}, \citenamefont {Pollmann},\ and\
  \citenamefont {Sondhi}}]{von2018operator}%
  \BibitemOpen
  \bibfield  {author} {\bibinfo {author} {\bibfnamefont {C.}~\bibnamefont
  {Von~Keyserlingk}}, \bibinfo {author} {\bibfnamefont {T.}~\bibnamefont
  {Rakovszky}}, \bibinfo {author} {\bibfnamefont {F.}~\bibnamefont {Pollmann}},
  \ and\ \bibinfo {author} {\bibfnamefont {S.~L.}\ \bibnamefont {Sondhi}},\
  }\href@noop {} {\bibfield  {journal} {\bibinfo  {journal} {Physical Review
  X}\ }\textbf {\bibinfo {volume} {8}},\ \bibinfo {pages} {021013} (\bibinfo
  {year} {2018})}\BibitemShut {NoStop}%
\bibitem [{\citenamefont {Nahum}\ \emph {et~al.}(2018)\citenamefont {Nahum},
  \citenamefont {Vijay},\ and\ \citenamefont {Haah}}]{nahum2018operator}%
  \BibitemOpen
  \bibfield  {author} {\bibinfo {author} {\bibfnamefont {A.}~\bibnamefont
  {Nahum}}, \bibinfo {author} {\bibfnamefont {S.}~\bibnamefont {Vijay}}, \ and\
  \bibinfo {author} {\bibfnamefont {J.}~\bibnamefont {Haah}},\ }\href@noop {}
  {\bibfield  {journal} {\bibinfo  {journal} {Physical Review X}\ }\textbf
  {\bibinfo {volume} {8}},\ \bibinfo {pages} {021014} (\bibinfo {year}
  {2018})}\BibitemShut {NoStop}%
\bibitem [{\citenamefont {Lin}\ and\ \citenamefont
  {Motrunich}(2018)}]{PhysRevB.97.144304}%
  \BibitemOpen
  \bibfield  {author} {\bibinfo {author} {\bibfnamefont {C.-J.}\ \bibnamefont
  {Lin}}\ and\ \bibinfo {author} {\bibfnamefont {O.~I.}\ \bibnamefont
  {Motrunich}},\ }\href {\doibase 10.1103/PhysRevB.97.144304} {\bibfield
  {journal} {\bibinfo  {journal} {Phys. Rev. B}\ }\textbf {\bibinfo {volume}
  {97}},\ \bibinfo {pages} {144304} (\bibinfo {year} {2018})}\BibitemShut
  {NoStop}%
\bibitem [{\citenamefont {Xu}\ and\ \citenamefont
  {Swingle}(2020)}]{xu2020accessing}%
  \BibitemOpen
  \bibfield  {author} {\bibinfo {author} {\bibfnamefont {S.}~\bibnamefont
  {Xu}}\ and\ \bibinfo {author} {\bibfnamefont {B.}~\bibnamefont {Swingle}},\
  }\href@noop {} {\bibfield  {journal} {\bibinfo  {journal} {Nature Physics}\
  }\textbf {\bibinfo {volume} {16}},\ \bibinfo {pages} {199} (\bibinfo {year}
  {2020})}\BibitemShut {NoStop}%
\bibitem [{\citenamefont {Gopalakrishnan}\ \emph {et~al.}(2018)\citenamefont
  {Gopalakrishnan}, \citenamefont {Huse}, \citenamefont {Khemani},\ and\
  \citenamefont {Vasseur}}]{gopalakrishnan2018hydrodynamics}%
  \BibitemOpen
  \bibfield  {author} {\bibinfo {author} {\bibfnamefont {S.}~\bibnamefont
  {Gopalakrishnan}}, \bibinfo {author} {\bibfnamefont {D.~A.}\ \bibnamefont
  {Huse}}, \bibinfo {author} {\bibfnamefont {V.}~\bibnamefont {Khemani}}, \
  and\ \bibinfo {author} {\bibfnamefont {R.}~\bibnamefont {Vasseur}},\
  }\href@noop {} {\bibfield  {journal} {\bibinfo  {journal} {Physical Review
  B}\ }\textbf {\bibinfo {volume} {98}},\ \bibinfo {pages} {220303} (\bibinfo
  {year} {2018})}\BibitemShut {NoStop}%
\bibitem [{\citenamefont {Swingle}\ \emph {et~al.}(2016)\citenamefont
  {Swingle}, \citenamefont {Bentsen}, \citenamefont {Schleier-Smith},\ and\
  \citenamefont {Hayden}}]{PhysRevA.94.040302}%
  \BibitemOpen
  \bibfield  {author} {\bibinfo {author} {\bibfnamefont {B.}~\bibnamefont
  {Swingle}}, \bibinfo {author} {\bibfnamefont {G.}~\bibnamefont {Bentsen}},
  \bibinfo {author} {\bibfnamefont {M.}~\bibnamefont {Schleier-Smith}}, \ and\
  \bibinfo {author} {\bibfnamefont {P.}~\bibnamefont {Hayden}},\ }\href
  {\doibase 10.1103/PhysRevA.94.040302} {\bibfield  {journal} {\bibinfo
  {journal} {Phys. Rev. A}\ }\textbf {\bibinfo {volume} {94}},\ \bibinfo
  {pages} {040302} (\bibinfo {year} {2016})}\BibitemShut {NoStop}%
\bibitem [{\citenamefont {Zhu}\ \emph {et~al.}(2016)\citenamefont {Zhu},
  \citenamefont {Hafezi},\ and\ \citenamefont {Grover}}]{zhu2016measurement}%
  \BibitemOpen
  \bibfield  {author} {\bibinfo {author} {\bibfnamefont {G.}~\bibnamefont
  {Zhu}}, \bibinfo {author} {\bibfnamefont {M.}~\bibnamefont {Hafezi}}, \ and\
  \bibinfo {author} {\bibfnamefont {T.}~\bibnamefont {Grover}},\ }\href@noop {}
  {\bibfield  {journal} {\bibinfo  {journal} {Physical Review A}\ }\textbf
  {\bibinfo {volume} {94}},\ \bibinfo {pages} {062329} (\bibinfo {year}
  {2016})}\BibitemShut {NoStop}%
\bibitem [{\citenamefont {G{\"a}rttner}\ \emph {et~al.}(2017)\citenamefont
  {G{\"a}rttner}, \citenamefont {Bohnet}, \citenamefont {Safavi-Naini},
  \citenamefont {Wall}, \citenamefont {Bollinger},\ and\ \citenamefont
  {Rey}}]{garttner2017measuring}%
  \BibitemOpen
  \bibfield  {author} {\bibinfo {author} {\bibfnamefont {M.}~\bibnamefont
  {G{\"a}rttner}}, \bibinfo {author} {\bibfnamefont {J.~G.}\ \bibnamefont
  {Bohnet}}, \bibinfo {author} {\bibfnamefont {A.}~\bibnamefont
  {Safavi-Naini}}, \bibinfo {author} {\bibfnamefont {M.~L.}\ \bibnamefont
  {Wall}}, \bibinfo {author} {\bibfnamefont {J.~J.}\ \bibnamefont {Bollinger}},
  \ and\ \bibinfo {author} {\bibfnamefont {A.~M.}\ \bibnamefont {Rey}},\
  }\href@noop {} {\bibfield  {journal} {\bibinfo  {journal} {Nature Physics}\
  }\textbf {\bibinfo {volume} {13}},\ \bibinfo {pages} {781} (\bibinfo {year}
  {2017})}\BibitemShut {NoStop}%
\bibitem [{\citenamefont {Yao}\ \emph {et~al.}(2016)\citenamefont {Yao},
  \citenamefont {Grusdt}, \citenamefont {Swingle}, \citenamefont {Lukin},
  \citenamefont {Stamper-Kurn}, \citenamefont {Moore},\ and\ \citenamefont
  {Demler}}]{yao2016interferometric}%
  \BibitemOpen
  \bibfield  {author} {\bibinfo {author} {\bibfnamefont {N.~Y.}\ \bibnamefont
  {Yao}}, \bibinfo {author} {\bibfnamefont {F.}~\bibnamefont {Grusdt}},
  \bibinfo {author} {\bibfnamefont {B.}~\bibnamefont {Swingle}}, \bibinfo
  {author} {\bibfnamefont {M.~D.}\ \bibnamefont {Lukin}}, \bibinfo {author}
  {\bibfnamefont {D.~M.}\ \bibnamefont {Stamper-Kurn}}, \bibinfo {author}
  {\bibfnamefont {J.~E.}\ \bibnamefont {Moore}}, \ and\ \bibinfo {author}
  {\bibfnamefont {E.~A.}\ \bibnamefont {Demler}},\ }\href@noop {} {\bibfield
  {journal} {\bibinfo  {journal} {arXiv preprint arXiv:1607.01801}\ } (\bibinfo
  {year} {2016})}\BibitemShut {NoStop}%
\bibitem [{\citenamefont {Wei}\ \emph {et~al.}(2019)\citenamefont {Wei},
  \citenamefont {Peng}, \citenamefont {Shtanko}, \citenamefont {Marvian},
  \citenamefont {Lloyd}, \citenamefont {Ramanathan},\ and\ \citenamefont
  {Cappellaro}}]{PhysRevLett.123.090605}%
  \BibitemOpen
  \bibfield  {author} {\bibinfo {author} {\bibfnamefont {K.~X.}\ \bibnamefont
  {Wei}}, \bibinfo {author} {\bibfnamefont {P.}~\bibnamefont {Peng}}, \bibinfo
  {author} {\bibfnamefont {O.}~\bibnamefont {Shtanko}}, \bibinfo {author}
  {\bibfnamefont {I.}~\bibnamefont {Marvian}}, \bibinfo {author} {\bibfnamefont
  {S.}~\bibnamefont {Lloyd}}, \bibinfo {author} {\bibfnamefont
  {C.}~\bibnamefont {Ramanathan}}, \ and\ \bibinfo {author} {\bibfnamefont
  {P.}~\bibnamefont {Cappellaro}},\ }\href {\doibase
  10.1103/PhysRevLett.123.090605} {\bibfield  {journal} {\bibinfo  {journal}
  {Phys. Rev. Lett.}\ }\textbf {\bibinfo {volume} {123}},\ \bibinfo {pages}
  {090605} (\bibinfo {year} {2019})}\BibitemShut {NoStop}%
\bibitem [{\citenamefont {{Mi}}\ \emph {et~al.}(2021)\citenamefont {{Mi}},
  \citenamefont {{Roushan}}, \citenamefont {{Quintana}}, \citenamefont
  {{Mandra}}, \citenamefont {{Marshall}}, \citenamefont {{Neill}},
  \citenamefont {{Arute}}, \citenamefont {{Arya}}, \citenamefont {{Atalaya}},
  \citenamefont {{Babbush}}, \citenamefont {{Bardin}}, \citenamefont
  {{Barends}}, \citenamefont {{Bengtsson}}, \citenamefont {{Boixo}},
  \citenamefont {{Bourassa}}, \citenamefont {{Broughton}}, \citenamefont
  {{Buckley}}, \citenamefont {{Buell}}, \citenamefont {{Burkett}},
  \citenamefont {{Bushnell}}, \citenamefont {{Chen}}, \citenamefont {{Chiaro}},
  \citenamefont {{Collins}}, \citenamefont {{Courtney}}, \citenamefont
  {{Demura}}, \citenamefont {{Derk}}, \citenamefont {{Dunsworth}},
  \citenamefont {{Eppens}}, \citenamefont {{Erickson}}, \citenamefont
  {{Farhi}}, \citenamefont {{Fowler}}, \citenamefont {{Foxen}}, \citenamefont
  {{Gidney}}, \citenamefont {{Giustina}}, \citenamefont {{Gross}},
  \citenamefont {{Harrigan}}, \citenamefont {{Harrington}}, \citenamefont
  {{Hilton}}, \citenamefont {{Ho}}, \citenamefont {{Hong}}, \citenamefont
  {{Huang}}, \citenamefont {{Huggins}}, \citenamefont {{Ioffe}}, \citenamefont
  {{Isakov}}, \citenamefont {{Jeffrey}}, \citenamefont {{Jiang}}, \citenamefont
  {{Jones}}, \citenamefont {{Kafri}}, \citenamefont {{Kelly}}, \citenamefont
  {{Kim}}, \citenamefont {{Kitaev}}, \citenamefont {{Klimov}}, \citenamefont
  {{Korotkov}}, \citenamefont {{Kostritsa}}, \citenamefont {{Landhuis}},
  \citenamefont {{Laptev}}, \citenamefont {{Lucero}}, \citenamefont {{Martin}},
  \citenamefont {{McClean}}, \citenamefont {{McCourt}}, \citenamefont
  {{McEwen}}, \citenamefont {{Megrant}}, \citenamefont {{Miao}}, \citenamefont
  {{Mohseni}}, \citenamefont {{Mruczkiewicz}}, \citenamefont {{Mutus}},
  \citenamefont {{Naaman}}, \citenamefont {{Neeley}}, \citenamefont {{Newman}},
  \citenamefont {{Yuezhen Niu}}, \citenamefont {{O'Brien}}, \citenamefont
  {{Opremcak}}, \citenamefont {{Ostby}}, \citenamefont {{Pato}}, \citenamefont
  {{Petukhov}}, \citenamefont {{Redd}}, \citenamefont {{Rubin}}, \citenamefont
  {{Sank}}, \citenamefont {{Satzinger}}, \citenamefont {{Shvarts}},
  \citenamefont {{Strain}}, \citenamefont {{Szalay}}, \citenamefont
  {{Trevithick}}, \citenamefont {{Villalonga}}, \citenamefont {{White}},
  \citenamefont {{Yao}}, \citenamefont {{Yeh}}, \citenamefont {{Zalcman}},
  \citenamefont {{Neven}}, \citenamefont {{Aleiner}}, \citenamefont
  {{Kechedzhi}}, \citenamefont {{Smelyanskiy}},\ and\ \citenamefont
  {{Chen}}}]{2021arXiv210108870M}%
  \BibitemOpen
  \bibfield  {author} {\bibinfo {author} {\bibfnamefont {X.}~\bibnamefont
  {{Mi}}}, \bibinfo {author} {\bibfnamefont {P.}~\bibnamefont {{Roushan}}},
  \bibinfo {author} {\bibfnamefont {C.}~\bibnamefont {{Quintana}}}, \bibinfo
  {author} {\bibfnamefont {S.}~\bibnamefont {{Mandra}}}, \bibinfo {author}
  {\bibfnamefont {J.}~\bibnamefont {{Marshall}}}, \bibinfo {author}
  {\bibfnamefont {C.}~\bibnamefont {{Neill}}}, \bibinfo {author} {\bibfnamefont
  {F.}~\bibnamefont {{Arute}}}, \bibinfo {author} {\bibfnamefont
  {K.}~\bibnamefont {{Arya}}}, \bibinfo {author} {\bibfnamefont
  {J.}~\bibnamefont {{Atalaya}}}, \bibinfo {author} {\bibfnamefont
  {R.}~\bibnamefont {{Babbush}}}, \bibinfo {author} {\bibfnamefont {J.~C.}\
  \bibnamefont {{Bardin}}}, \bibinfo {author} {\bibfnamefont {R.}~\bibnamefont
  {{Barends}}}, \bibinfo {author} {\bibfnamefont {A.}~\bibnamefont
  {{Bengtsson}}}, \bibinfo {author} {\bibfnamefont {S.}~\bibnamefont
  {{Boixo}}}, \bibinfo {author} {\bibfnamefont {A.}~\bibnamefont {{Bourassa}}},
  \bibinfo {author} {\bibfnamefont {M.}~\bibnamefont {{Broughton}}}, \bibinfo
  {author} {\bibfnamefont {B.~B.}\ \bibnamefont {{Buckley}}}, \bibinfo {author}
  {\bibfnamefont {D.~A.}\ \bibnamefont {{Buell}}}, \bibinfo {author}
  {\bibfnamefont {B.}~\bibnamefont {{Burkett}}}, \bibinfo {author}
  {\bibfnamefont {N.}~\bibnamefont {{Bushnell}}}, \bibinfo {author}
  {\bibfnamefont {Z.}~\bibnamefont {{Chen}}}, \bibinfo {author} {\bibfnamefont
  {B.}~\bibnamefont {{Chiaro}}}, \bibinfo {author} {\bibfnamefont
  {R.}~\bibnamefont {{Collins}}}, \bibinfo {author} {\bibfnamefont
  {W.}~\bibnamefont {{Courtney}}}, \bibinfo {author} {\bibfnamefont
  {S.}~\bibnamefont {{Demura}}}, \bibinfo {author} {\bibfnamefont {A.~R.}\
  \bibnamefont {{Derk}}}, \bibinfo {author} {\bibfnamefont {A.}~\bibnamefont
  {{Dunsworth}}}, \bibinfo {author} {\bibfnamefont {D.}~\bibnamefont
  {{Eppens}}}, \bibinfo {author} {\bibfnamefont {C.}~\bibnamefont
  {{Erickson}}}, \bibinfo {author} {\bibfnamefont {E.}~\bibnamefont {{Farhi}}},
  \bibinfo {author} {\bibfnamefont {A.~G.}\ \bibnamefont {{Fowler}}}, \bibinfo
  {author} {\bibfnamefont {B.}~\bibnamefont {{Foxen}}}, \bibinfo {author}
  {\bibfnamefont {C.}~\bibnamefont {{Gidney}}}, \bibinfo {author}
  {\bibfnamefont {M.}~\bibnamefont {{Giustina}}}, \bibinfo {author}
  {\bibfnamefont {J.~A.}\ \bibnamefont {{Gross}}}, \bibinfo {author}
  {\bibfnamefont {M.~P.}\ \bibnamefont {{Harrigan}}}, \bibinfo {author}
  {\bibfnamefont {S.~D.}\ \bibnamefont {{Harrington}}}, \bibinfo {author}
  {\bibfnamefont {J.}~\bibnamefont {{Hilton}}}, \bibinfo {author}
  {\bibfnamefont {A.}~\bibnamefont {{Ho}}}, \bibinfo {author} {\bibfnamefont
  {S.}~\bibnamefont {{Hong}}}, \bibinfo {author} {\bibfnamefont
  {T.}~\bibnamefont {{Huang}}}, \bibinfo {author} {\bibfnamefont {W.~J.}\
  \bibnamefont {{Huggins}}}, \bibinfo {author} {\bibfnamefont {L.~B.}\
  \bibnamefont {{Ioffe}}}, \bibinfo {author} {\bibfnamefont {S.~V.}\
  \bibnamefont {{Isakov}}}, \bibinfo {author} {\bibfnamefont {E.}~\bibnamefont
  {{Jeffrey}}}, \bibinfo {author} {\bibfnamefont {Z.}~\bibnamefont {{Jiang}}},
  \bibinfo {author} {\bibfnamefont {C.}~\bibnamefont {{Jones}}}, \bibinfo
  {author} {\bibfnamefont {D.}~\bibnamefont {{Kafri}}}, \bibinfo {author}
  {\bibfnamefont {J.}~\bibnamefont {{Kelly}}}, \bibinfo {author} {\bibfnamefont
  {S.}~\bibnamefont {{Kim}}}, \bibinfo {author} {\bibfnamefont
  {A.}~\bibnamefont {{Kitaev}}}, \bibinfo {author} {\bibfnamefont {P.~V.}\
  \bibnamefont {{Klimov}}}, \bibinfo {author} {\bibfnamefont {A.~N.}\
  \bibnamefont {{Korotkov}}}, \bibinfo {author} {\bibfnamefont
  {F.}~\bibnamefont {{Kostritsa}}}, \bibinfo {author} {\bibfnamefont
  {D.}~\bibnamefont {{Landhuis}}}, \bibinfo {author} {\bibfnamefont
  {P.}~\bibnamefont {{Laptev}}}, \bibinfo {author} {\bibfnamefont
  {E.}~\bibnamefont {{Lucero}}}, \bibinfo {author} {\bibfnamefont
  {O.}~\bibnamefont {{Martin}}}, \bibinfo {author} {\bibfnamefont {J.~R.}\
  \bibnamefont {{McClean}}}, \bibinfo {author} {\bibfnamefont {T.}~\bibnamefont
  {{McCourt}}}, \bibinfo {author} {\bibfnamefont {M.}~\bibnamefont {{McEwen}}},
  \bibinfo {author} {\bibfnamefont {A.}~\bibnamefont {{Megrant}}}, \bibinfo
  {author} {\bibfnamefont {K.~C.}\ \bibnamefont {{Miao}}}, \bibinfo {author}
  {\bibfnamefont {M.}~\bibnamefont {{Mohseni}}}, \bibinfo {author}
  {\bibfnamefont {W.}~\bibnamefont {{Mruczkiewicz}}}, \bibinfo {author}
  {\bibfnamefont {J.}~\bibnamefont {{Mutus}}}, \bibinfo {author} {\bibfnamefont
  {O.}~\bibnamefont {{Naaman}}}, \bibinfo {author} {\bibfnamefont
  {M.}~\bibnamefont {{Neeley}}}, \bibinfo {author} {\bibfnamefont
  {M.}~\bibnamefont {{Newman}}}, \bibinfo {author} {\bibfnamefont
  {M.}~\bibnamefont {{Yuezhen Niu}}}, \bibinfo {author} {\bibfnamefont {T.~E.}\
  \bibnamefont {{O'Brien}}}, \bibinfo {author} {\bibfnamefont {A.}~\bibnamefont
  {{Opremcak}}}, \bibinfo {author} {\bibfnamefont {E.}~\bibnamefont {{Ostby}}},
  \bibinfo {author} {\bibfnamefont {B.}~\bibnamefont {{Pato}}}, \bibinfo
  {author} {\bibfnamefont {A.}~\bibnamefont {{Petukhov}}}, \bibinfo {author}
  {\bibfnamefont {N.}~\bibnamefont {{Redd}}}, \bibinfo {author} {\bibfnamefont
  {N.~C.}\ \bibnamefont {{Rubin}}}, \bibinfo {author} {\bibfnamefont
  {D.}~\bibnamefont {{Sank}}}, \bibinfo {author} {\bibfnamefont {K.~J.}\
  \bibnamefont {{Satzinger}}}, \bibinfo {author} {\bibfnamefont
  {V.}~\bibnamefont {{Shvarts}}}, \bibinfo {author} {\bibfnamefont
  {D.}~\bibnamefont {{Strain}}}, \bibinfo {author} {\bibfnamefont
  {M.}~\bibnamefont {{Szalay}}}, \bibinfo {author} {\bibfnamefont {M.~D.}\
  \bibnamefont {{Trevithick}}}, \bibinfo {author} {\bibfnamefont
  {B.}~\bibnamefont {{Villalonga}}}, \bibinfo {author} {\bibfnamefont
  {T.}~\bibnamefont {{White}}}, \bibinfo {author} {\bibfnamefont {Z.~J.}\
  \bibnamefont {{Yao}}}, \bibinfo {author} {\bibfnamefont {P.}~\bibnamefont
  {{Yeh}}}, \bibinfo {author} {\bibfnamefont {A.}~\bibnamefont {{Zalcman}}},
  \bibinfo {author} {\bibfnamefont {H.}~\bibnamefont {{Neven}}}, \bibinfo
  {author} {\bibfnamefont {I.}~\bibnamefont {{Aleiner}}}, \bibinfo {author}
  {\bibfnamefont {K.}~\bibnamefont {{Kechedzhi}}}, \bibinfo {author}
  {\bibfnamefont {V.}~\bibnamefont {{Smelyanskiy}}}, \ and\ \bibinfo {author}
  {\bibfnamefont {Y.}~\bibnamefont {{Chen}}},\ }\href@noop {} {\bibfield
  {journal} {\bibinfo  {journal} {arXiv e-prints}\ ,\ \bibinfo {eid}
  {arXiv:2101.08870}} (\bibinfo {year} {2021})},\ \Eprint
  {http://arxiv.org/abs/2101.08870} {arXiv:2101.08870 [quant-ph]} \BibitemShut
  {NoStop}%
\bibitem [{\citenamefont {Prosen}\ and\ \citenamefont {\ifmmode \check{Z}\else
  \v{Z}\fi{}nidari\ifmmode~\check{c}\else
  \v{c}\fi{}}(2007)}]{PhysRevE.75.015202}%
  \BibitemOpen
  \bibfield  {author} {\bibinfo {author} {\bibfnamefont {T.~c.~v.}\
  \bibnamefont {Prosen}}\ and\ \bibinfo {author} {\bibfnamefont
  {M.}~\bibnamefont {\ifmmode \check{Z}\else
  \v{Z}\fi{}nidari\ifmmode~\check{c}\else \v{c}\fi{}}},\ }\href {\doibase
  10.1103/PhysRevE.75.015202} {\bibfield  {journal} {\bibinfo  {journal} {Phys.
  Rev. E}\ }\textbf {\bibinfo {volume} {75}},\ \bibinfo {pages} {015202}
  (\bibinfo {year} {2007})}\BibitemShut {NoStop}%
\bibitem [{\citenamefont {Alba}\ \emph {et~al.}(2019)\citenamefont {Alba},
  \citenamefont {Dubail},\ and\ \citenamefont
  {Medenjak}}]{PhysRevLett.122.250603}%
  \BibitemOpen
  \bibfield  {author} {\bibinfo {author} {\bibfnamefont {V.}~\bibnamefont
  {Alba}}, \bibinfo {author} {\bibfnamefont {J.}~\bibnamefont {Dubail}}, \ and\
  \bibinfo {author} {\bibfnamefont {M.}~\bibnamefont {Medenjak}},\ }\href
  {\doibase 10.1103/PhysRevLett.122.250603} {\bibfield  {journal} {\bibinfo
  {journal} {Phys. Rev. Lett.}\ }\textbf {\bibinfo {volume} {122}},\ \bibinfo
  {pages} {250603} (\bibinfo {year} {2019})}\BibitemShut {NoStop}%
\bibitem [{\citenamefont {Calabrese}\ and\ \citenamefont
  {Cardy}(2006)}]{PhysRevLett.96.136801}%
  \BibitemOpen
  \bibfield  {author} {\bibinfo {author} {\bibfnamefont {P.}~\bibnamefont
  {Calabrese}}\ and\ \bibinfo {author} {\bibfnamefont {J.}~\bibnamefont
  {Cardy}},\ }\href {\doibase 10.1103/PhysRevLett.96.136801} {\bibfield
  {journal} {\bibinfo  {journal} {Phys. Rev. Lett.}\ }\textbf {\bibinfo
  {volume} {96}},\ \bibinfo {pages} {136801} (\bibinfo {year}
  {2006})}\BibitemShut {NoStop}%
\bibitem [{\citenamefont {Prosen}(2011)}]{PhysRevLett.106.217206}%
  \BibitemOpen
  \bibfield  {author} {\bibinfo {author} {\bibfnamefont {T.~c.~v.}\
  \bibnamefont {Prosen}},\ }\href {\doibase 10.1103/PhysRevLett.106.217206}
  {\bibfield  {journal} {\bibinfo  {journal} {Phys. Rev. Lett.}\ }\textbf
  {\bibinfo {volume} {106}},\ \bibinfo {pages} {217206} (\bibinfo {year}
  {2011})}\BibitemShut {NoStop}%
\bibitem [{\citenamefont {Caux}\ and\ \citenamefont
  {Essler}(2013)}]{PhysRevLett.110.257203}%
  \BibitemOpen
  \bibfield  {author} {\bibinfo {author} {\bibfnamefont {J.-S.}\ \bibnamefont
  {Caux}}\ and\ \bibinfo {author} {\bibfnamefont {F.~H.~L.}\ \bibnamefont
  {Essler}},\ }\href {\doibase 10.1103/PhysRevLett.110.257203} {\bibfield
  {journal} {\bibinfo  {journal} {Phys. Rev. Lett.}\ }\textbf {\bibinfo
  {volume} {110}},\ \bibinfo {pages} {257203} (\bibinfo {year}
  {2013})}\BibitemShut {NoStop}%
\bibitem [{\citenamefont {Wouters}\ \emph {et~al.}(2014)\citenamefont
  {Wouters}, \citenamefont {De~Nardis}, \citenamefont {Brockmann},
  \citenamefont {Fioretto}, \citenamefont {Rigol},\ and\ \citenamefont
  {Caux}}]{PhysRevLett.113.117202}%
  \BibitemOpen
  \bibfield  {author} {\bibinfo {author} {\bibfnamefont {B.}~\bibnamefont
  {Wouters}}, \bibinfo {author} {\bibfnamefont {J.}~\bibnamefont {De~Nardis}},
  \bibinfo {author} {\bibfnamefont {M.}~\bibnamefont {Brockmann}}, \bibinfo
  {author} {\bibfnamefont {D.}~\bibnamefont {Fioretto}}, \bibinfo {author}
  {\bibfnamefont {M.}~\bibnamefont {Rigol}}, \ and\ \bibinfo {author}
  {\bibfnamefont {J.-S.}\ \bibnamefont {Caux}},\ }\href {\doibase
  10.1103/PhysRevLett.113.117202} {\bibfield  {journal} {\bibinfo  {journal}
  {Phys. Rev. Lett.}\ }\textbf {\bibinfo {volume} {113}},\ \bibinfo {pages}
  {117202} (\bibinfo {year} {2014})}\BibitemShut {NoStop}%
\bibitem [{\citenamefont {Ilievski}\ \emph {et~al.}(2015)\citenamefont
  {Ilievski}, \citenamefont {De~Nardis}, \citenamefont {Wouters}, \citenamefont
  {Caux}, \citenamefont {Essler},\ and\ \citenamefont
  {Prosen}}]{PhysRevLett.115.157201}%
  \BibitemOpen
  \bibfield  {author} {\bibinfo {author} {\bibfnamefont {E.}~\bibnamefont
  {Ilievski}}, \bibinfo {author} {\bibfnamefont {J.}~\bibnamefont {De~Nardis}},
  \bibinfo {author} {\bibfnamefont {B.}~\bibnamefont {Wouters}}, \bibinfo
  {author} {\bibfnamefont {J.-S.}\ \bibnamefont {Caux}}, \bibinfo {author}
  {\bibfnamefont {F.~H.~L.}\ \bibnamefont {Essler}}, \ and\ \bibinfo {author}
  {\bibfnamefont {T.}~\bibnamefont {Prosen}},\ }\href {\doibase
  10.1103/PhysRevLett.115.157201} {\bibfield  {journal} {\bibinfo  {journal}
  {Phys. Rev. Lett.}\ }\textbf {\bibinfo {volume} {115}},\ \bibinfo {pages}
  {157201} (\bibinfo {year} {2015})}\BibitemShut {NoStop}%
\bibitem [{\citenamefont {Ilievski}\ \emph {et~al.}(2016)\citenamefont
  {Ilievski}, \citenamefont {Medenjak}, \citenamefont {Prosen},\ and\
  \citenamefont {Zadnik}}]{ilievski2016quasilocal}%
  \BibitemOpen
  \bibfield  {author} {\bibinfo {author} {\bibfnamefont {E.}~\bibnamefont
  {Ilievski}}, \bibinfo {author} {\bibfnamefont {M.}~\bibnamefont {Medenjak}},
  \bibinfo {author} {\bibfnamefont {T.}~\bibnamefont {Prosen}}, \ and\ \bibinfo
  {author} {\bibfnamefont {L.}~\bibnamefont {Zadnik}},\ }\href@noop {}
  {\bibfield  {journal} {\bibinfo  {journal} {Journal of Statistical Mechanics:
  Theory and Experiment}\ }\textbf {\bibinfo {volume} {2016}},\ \bibinfo
  {pages} {064008} (\bibinfo {year} {2016})}\BibitemShut {NoStop}%
\bibitem [{\citenamefont {Vasseur}\ and\ \citenamefont
  {Moore}(2016)}]{vasseur2016nonequilibrium}%
  \BibitemOpen
  \bibfield  {author} {\bibinfo {author} {\bibfnamefont {R.}~\bibnamefont
  {Vasseur}}\ and\ \bibinfo {author} {\bibfnamefont {J.~E.}\ \bibnamefont
  {Moore}},\ }\href@noop {} {\bibfield  {journal} {\bibinfo  {journal} {Journal
  of Statistical Mechanics: Theory and Experiment}\ }\textbf {\bibinfo {volume}
  {2016}},\ \bibinfo {pages} {064010} (\bibinfo {year} {2016})}\BibitemShut
  {NoStop}%
\bibitem [{\citenamefont {Fagotti}\ \emph {et~al.}(2014)\citenamefont
  {Fagotti}, \citenamefont {Collura}, \citenamefont {Essler},\ and\
  \citenamefont {Calabrese}}]{PhysRevB.89.125101}%
  \BibitemOpen
  \bibfield  {author} {\bibinfo {author} {\bibfnamefont {M.}~\bibnamefont
  {Fagotti}}, \bibinfo {author} {\bibfnamefont {M.}~\bibnamefont {Collura}},
  \bibinfo {author} {\bibfnamefont {F.~H.~L.}\ \bibnamefont {Essler}}, \ and\
  \bibinfo {author} {\bibfnamefont {P.}~\bibnamefont {Calabrese}},\ }\href
  {\doibase 10.1103/PhysRevB.89.125101} {\bibfield  {journal} {\bibinfo
  {journal} {Phys. Rev. B}\ }\textbf {\bibinfo {volume} {89}},\ \bibinfo
  {pages} {125101} (\bibinfo {year} {2014})}\BibitemShut {NoStop}%
\bibitem [{\citenamefont {Alba}\ and\ \citenamefont
  {Calabrese}(2017{\natexlab{a}})}]{alba2017entanglement}%
  \BibitemOpen
  \bibfield  {author} {\bibinfo {author} {\bibfnamefont {V.}~\bibnamefont
  {Alba}}\ and\ \bibinfo {author} {\bibfnamefont {P.}~\bibnamefont
  {Calabrese}},\ }\href@noop {} {\bibfield  {journal} {\bibinfo  {journal}
  {Proceedings of the National Academy of Sciences}\ }\textbf {\bibinfo
  {volume} {114}},\ \bibinfo {pages} {7947} (\bibinfo {year}
  {2017}{\natexlab{a}})}\BibitemShut {NoStop}%
\bibitem [{\citenamefont {Platini}\ and\ \citenamefont
  {Karevski}(2005)}]{Platini2005}%
  \BibitemOpen
  \bibfield  {author} {\bibinfo {author} {\bibfnamefont {T.}~\bibnamefont
  {Platini}}\ and\ \bibinfo {author} {\bibfnamefont {D.}~\bibnamefont
  {Karevski}},\ }\href {\doibase 10.1140/epjb/e2005-00402-2} {\bibfield
  {journal} {\bibinfo  {journal} {The European Physical Journal B - Condensed
  Matter and Complex Systems}\ }\textbf {\bibinfo {volume} {48}},\ \bibinfo
  {pages} {225} (\bibinfo {year} {2005})}\BibitemShut {NoStop}%
\bibitem [{\citenamefont {Collura}\ \emph {et~al.}(2018)\citenamefont
  {Collura}, \citenamefont {De~Luca},\ and\ \citenamefont
  {Viti}}]{PhysRevB.97.081111}%
  \BibitemOpen
  \bibfield  {author} {\bibinfo {author} {\bibfnamefont {M.}~\bibnamefont
  {Collura}}, \bibinfo {author} {\bibfnamefont {A.}~\bibnamefont {De~Luca}}, \
  and\ \bibinfo {author} {\bibfnamefont {J.}~\bibnamefont {Viti}},\ }\href
  {\doibase 10.1103/PhysRevB.97.081111} {\bibfield  {journal} {\bibinfo
  {journal} {Phys. Rev. B}\ }\textbf {\bibinfo {volume} {97}},\ \bibinfo
  {pages} {081111} (\bibinfo {year} {2018})}\BibitemShut {NoStop}%
\bibitem [{\citenamefont {Khemani}\ \emph
  {et~al.}(2018{\natexlab{b}})\citenamefont {Khemani}, \citenamefont {Huse},\
  and\ \citenamefont {Nahum}}]{2018arXiv180305902K}%
  \BibitemOpen
  \bibfield  {author} {\bibinfo {author} {\bibfnamefont {V.}~\bibnamefont
  {Khemani}}, \bibinfo {author} {\bibfnamefont {D.~A.}\ \bibnamefont {Huse}}, \
  and\ \bibinfo {author} {\bibfnamefont {A.}~\bibnamefont {Nahum}},\ }\href
  {\doibase 10.1103/PhysRevB.98.144304} {\bibfield  {journal} {\bibinfo
  {journal} {Phys. Rev. B}\ }\textbf {\bibinfo {volume} {98}},\ \bibinfo
  {pages} {144304} (\bibinfo {year} {2018}{\natexlab{b}})}\BibitemShut
  {NoStop}%
\bibitem [{\citenamefont {Fagotti}(2017)}]{PhysRevB.96.220302}%
  \BibitemOpen
  \bibfield  {author} {\bibinfo {author} {\bibfnamefont {M.}~\bibnamefont
  {Fagotti}},\ }\href {\doibase 10.1103/PhysRevB.96.220302} {\bibfield
  {journal} {\bibinfo  {journal} {Phys. Rev. B}\ }\textbf {\bibinfo {volume}
  {96}},\ \bibinfo {pages} {220302} (\bibinfo {year} {2017})}\BibitemShut
  {NoStop}%
\bibitem [{\citenamefont {Nardis}\ \emph {et~al.}(2019)\citenamefont {Nardis},
  \citenamefont {Bernard},\ and\ \citenamefont {Doyon}}]{de2019diffusion}%
  \BibitemOpen
  \bibfield  {author} {\bibinfo {author} {\bibfnamefont {J.~D.}\ \bibnamefont
  {Nardis}}, \bibinfo {author} {\bibfnamefont {D.}~\bibnamefont {Bernard}}, \
  and\ \bibinfo {author} {\bibfnamefont {B.}~\bibnamefont {Doyon}},\ }\href
  {\doibase 10.21468/SciPostPhys.6.4.049} {\bibfield  {journal} {\bibinfo
  {journal} {SciPost Phys.}\ }\textbf {\bibinfo {volume} {6}},\ \bibinfo
  {pages} {49} (\bibinfo {year} {2019})}\BibitemShut {NoStop}%
\bibitem [{\citenamefont {Gopalakrishnan}\ and\ \citenamefont
  {Zakirov}(2018)}]{Gopalakrishnan_2018}%
  \BibitemOpen
  \bibfield  {author} {\bibinfo {author} {\bibfnamefont {S.}~\bibnamefont
  {Gopalakrishnan}}\ and\ \bibinfo {author} {\bibfnamefont {B.}~\bibnamefont
  {Zakirov}},\ }\href {\doibase 10.1088/2058-9565/aad759} {\bibfield  {journal}
  {\bibinfo  {journal} {Quantum Science and Technology}\ }\textbf {\bibinfo
  {volume} {3}},\ \bibinfo {pages} {044004} (\bibinfo {year}
  {2018})}\BibitemShut {NoStop}%
\bibitem [{\citenamefont {Gopalakrishnan}(2018)}]{PhysRevB.98.060302}%
  \BibitemOpen
  \bibfield  {author} {\bibinfo {author} {\bibfnamefont {S.}~\bibnamefont
  {Gopalakrishnan}},\ }\href {\doibase 10.1103/PhysRevB.98.060302} {\bibfield
  {journal} {\bibinfo  {journal} {Phys. Rev. B}\ }\textbf {\bibinfo {volume}
  {98}},\ \bibinfo {pages} {060302} (\bibinfo {year} {2018})}\BibitemShut
  {NoStop}%
\bibitem [{\citenamefont {H\'emery}\ \emph {et~al.}(2019)\citenamefont
  {H\'emery}, \citenamefont {Pollmann},\ and\ \citenamefont
  {Luitz}}]{PhysRevB.100.104303}%
  \BibitemOpen
  \bibfield  {author} {\bibinfo {author} {\bibfnamefont {K.}~\bibnamefont
  {H\'emery}}, \bibinfo {author} {\bibfnamefont {F.}~\bibnamefont {Pollmann}},
  \ and\ \bibinfo {author} {\bibfnamefont {D.~J.}\ \bibnamefont {Luitz}},\
  }\href {\doibase 10.1103/PhysRevB.100.104303} {\bibfield  {journal} {\bibinfo
   {journal} {Phys. Rev. B}\ }\textbf {\bibinfo {volume} {100}},\ \bibinfo
  {pages} {104303} (\bibinfo {year} {2019})}\BibitemShut {NoStop}%
\bibitem [{\citenamefont {Claeys}\ and\ \citenamefont
  {Lamacraft}(2020)}]{PhysRevResearch.2.033032}%
  \BibitemOpen
  \bibfield  {author} {\bibinfo {author} {\bibfnamefont {P.~W.}\ \bibnamefont
  {Claeys}}\ and\ \bibinfo {author} {\bibfnamefont {A.}~\bibnamefont
  {Lamacraft}},\ }\href {\doibase 10.1103/PhysRevResearch.2.033032} {\bibfield
  {journal} {\bibinfo  {journal} {Phys. Rev. Research}\ }\textbf {\bibinfo
  {volume} {2}},\ \bibinfo {pages} {033032} (\bibinfo {year}
  {2020})}\BibitemShut {NoStop}%
\bibitem [{\citenamefont {Bertini}\ \emph {et~al.}(2019)\citenamefont
  {Bertini}, \citenamefont {Kos},\ and\ \citenamefont
  {Prosen}}]{PhysRevLett.123.210601}%
  \BibitemOpen
  \bibfield  {author} {\bibinfo {author} {\bibfnamefont {B.}~\bibnamefont
  {Bertini}}, \bibinfo {author} {\bibfnamefont {P.}~\bibnamefont {Kos}}, \ and\
  \bibinfo {author} {\bibfnamefont {T.~c.~v.}\ \bibnamefont {Prosen}},\ }\href
  {\doibase 10.1103/PhysRevLett.123.210601} {\bibfield  {journal} {\bibinfo
  {journal} {Phys. Rev. Lett.}\ }\textbf {\bibinfo {volume} {123}},\ \bibinfo
  {pages} {210601} (\bibinfo {year} {2019})}\BibitemShut {NoStop}%
\bibitem [{\citenamefont {Bertini}\ \emph {et~al.}(2020)\citenamefont
  {Bertini}, \citenamefont {Kos},\ and\ \citenamefont
  {Prosen}}]{bertini2020operator}%
  \BibitemOpen
  \bibfield  {author} {\bibinfo {author} {\bibfnamefont {B.}~\bibnamefont
  {Bertini}}, \bibinfo {author} {\bibfnamefont {P.}~\bibnamefont {Kos}}, \ and\
  \bibinfo {author} {\bibfnamefont {T.}~\bibnamefont {Prosen}},\ }\href@noop {}
  {\bibfield  {journal} {\bibinfo  {journal} {SciPost Phys}\ }\textbf {\bibinfo
  {volume} {8}},\ \bibinfo {pages} {067} (\bibinfo {year} {2020})}\BibitemShut
  {NoStop}%
\bibitem [{\citenamefont {Gopalakrishnan}\ and\ \citenamefont
  {Lamacraft}(2019)}]{PhysRevB.100.064309}%
  \BibitemOpen
  \bibfield  {author} {\bibinfo {author} {\bibfnamefont {S.}~\bibnamefont
  {Gopalakrishnan}}\ and\ \bibinfo {author} {\bibfnamefont {A.}~\bibnamefont
  {Lamacraft}},\ }\href {\doibase 10.1103/PhysRevB.100.064309} {\bibfield
  {journal} {\bibinfo  {journal} {Phys. Rev. B}\ }\textbf {\bibinfo {volume}
  {100}},\ \bibinfo {pages} {064309} (\bibinfo {year} {2019})}\BibitemShut
  {NoStop}%
\bibitem [{\citenamefont {Paeckel}\ \emph {et~al.}(2019)\citenamefont
  {Paeckel}, \citenamefont {K{\"o}hler}, \citenamefont {Swoboda}, \citenamefont
  {Manmana}, \citenamefont {Schollw{\"o}ck},\ and\ \citenamefont
  {Hubig}}]{paeckel2019time}%
  \BibitemOpen
  \bibfield  {author} {\bibinfo {author} {\bibfnamefont {S.}~\bibnamefont
  {Paeckel}}, \bibinfo {author} {\bibfnamefont {T.}~\bibnamefont {K{\"o}hler}},
  \bibinfo {author} {\bibfnamefont {A.}~\bibnamefont {Swoboda}}, \bibinfo
  {author} {\bibfnamefont {S.~R.}\ \bibnamefont {Manmana}}, \bibinfo {author}
  {\bibfnamefont {U.}~\bibnamefont {Schollw{\"o}ck}}, \ and\ \bibinfo {author}
  {\bibfnamefont {C.}~\bibnamefont {Hubig}},\ }\href@noop {} {\bibfield
  {journal} {\bibinfo  {journal} {Annals of Physics}\ }\textbf {\bibinfo
  {volume} {411}},\ \bibinfo {pages} {167998} (\bibinfo {year}
  {2019})}\BibitemShut {NoStop}%
\bibitem [{\citenamefont {Vidal}(2003)}]{PhysRevLett.91.147902}%
  \BibitemOpen
  \bibfield  {author} {\bibinfo {author} {\bibfnamefont {G.}~\bibnamefont
  {Vidal}},\ }\href {\doibase 10.1103/PhysRevLett.91.147902} {\bibfield
  {journal} {\bibinfo  {journal} {Phys. Rev. Lett.}\ }\textbf {\bibinfo
  {volume} {91}},\ \bibinfo {pages} {147902} (\bibinfo {year}
  {2003})}\BibitemShut {NoStop}%
\bibitem [{\citenamefont {Vidal}(2004)}]{PhysRevLett.93.040502}%
  \BibitemOpen
  \bibfield  {author} {\bibinfo {author} {\bibfnamefont {G.}~\bibnamefont
  {Vidal}},\ }\href {\doibase 10.1103/PhysRevLett.93.040502} {\bibfield
  {journal} {\bibinfo  {journal} {Phys. Rev. Lett.}\ }\textbf {\bibinfo
  {volume} {93}},\ \bibinfo {pages} {040502} (\bibinfo {year}
  {2004})}\BibitemShut {NoStop}%
\bibitem [{\citenamefont {Fishman}\ \emph {et~al.}(2020)\citenamefont
  {Fishman}, \citenamefont {White},\ and\ \citenamefont
  {Stoudenmire}}]{fishman2020itensor}%
  \BibitemOpen
  \bibfield  {author} {\bibinfo {author} {\bibfnamefont {M.}~\bibnamefont
  {Fishman}}, \bibinfo {author} {\bibfnamefont {S.~R.}\ \bibnamefont {White}},
  \ and\ \bibinfo {author} {\bibfnamefont {E.~M.}\ \bibnamefont
  {Stoudenmire}},\ }\href@noop {} {\enquote {\bibinfo {title} {The itensor
  software library for tensor network calculations},}\ } (\bibinfo {year}
  {2020}),\ \Eprint {http://arxiv.org/abs/2007.14822} {arXiv:2007.14822
  [cs.MS]} \BibitemShut {NoStop}%
\bibitem [{\citenamefont {De~Nardis}\ \emph {et~al.}(2018)\citenamefont
  {De~Nardis}, \citenamefont {Bernard},\ and\ \citenamefont
  {Doyon}}]{PhysRevLett.121.160603}%
  \BibitemOpen
  \bibfield  {author} {\bibinfo {author} {\bibfnamefont {J.}~\bibnamefont
  {De~Nardis}}, \bibinfo {author} {\bibfnamefont {D.}~\bibnamefont {Bernard}},
  \ and\ \bibinfo {author} {\bibfnamefont {B.}~\bibnamefont {Doyon}},\ }\href
  {\doibase 10.1103/PhysRevLett.121.160603} {\bibfield  {journal} {\bibinfo
  {journal} {Phys. Rev. Lett.}\ }\textbf {\bibinfo {volume} {121}},\ \bibinfo
  {pages} {160603} (\bibinfo {year} {2018})}\BibitemShut {NoStop}%
\bibitem [{\citenamefont {Hunyadi}\ \emph {et~al.}(2004)\citenamefont
  {Hunyadi}, \citenamefont {R\'acz},\ and\ \citenamefont
  {Sasv\'ari}}]{PhysRevE.69.066103}%
  \BibitemOpen
  \bibfield  {author} {\bibinfo {author} {\bibfnamefont {V.}~\bibnamefont
  {Hunyadi}}, \bibinfo {author} {\bibfnamefont {Z.}~\bibnamefont {R\'acz}}, \
  and\ \bibinfo {author} {\bibfnamefont {L.}~\bibnamefont {Sasv\'ari}},\ }\href
  {\doibase 10.1103/PhysRevE.69.066103} {\bibfield  {journal} {\bibinfo
  {journal} {Phys. Rev. E}\ }\textbf {\bibinfo {volume} {69}},\ \bibinfo
  {pages} {066103} (\bibinfo {year} {2004})}\BibitemShut {NoStop}%
\bibitem [{\citenamefont {Ljubotina}\ \emph {et~al.}(2017)\citenamefont
  {Ljubotina}, \citenamefont {{\v{Z}}nidari{\v{c}}},\ and\ \citenamefont
  {Prosen}}]{ljubotina2017spin}%
  \BibitemOpen
  \bibfield  {author} {\bibinfo {author} {\bibfnamefont {M.}~\bibnamefont
  {Ljubotina}}, \bibinfo {author} {\bibfnamefont {M.}~\bibnamefont
  {{\v{Z}}nidari{\v{c}}}}, \ and\ \bibinfo {author} {\bibfnamefont
  {T.}~\bibnamefont {Prosen}},\ }\href@noop {} {\bibfield  {journal} {\bibinfo
  {journal} {Nature communications}\ }\textbf {\bibinfo {volume} {8}},\
  \bibinfo {pages} {1} (\bibinfo {year} {2017})}\BibitemShut {NoStop}%
\bibitem [{\citenamefont {Bertini}\ \emph {et~al.}(2021)\citenamefont
  {Bertini}, \citenamefont {Heidrich-Meisner}, \citenamefont {Karrasch},
  \citenamefont {Prosen}, \citenamefont {Steinigeweg},\ and\ \citenamefont
  {{\v{Z}}nidari{\v{c}}}}]{bertini2021finite}%
  \BibitemOpen
  \bibfield  {author} {\bibinfo {author} {\bibfnamefont {B.}~\bibnamefont
  {Bertini}}, \bibinfo {author} {\bibfnamefont {F.}~\bibnamefont
  {Heidrich-Meisner}}, \bibinfo {author} {\bibfnamefont {C.}~\bibnamefont
  {Karrasch}}, \bibinfo {author} {\bibfnamefont {T.}~\bibnamefont {Prosen}},
  \bibinfo {author} {\bibfnamefont {R.}~\bibnamefont {Steinigeweg}}, \ and\
  \bibinfo {author} {\bibfnamefont {M.}~\bibnamefont {{\v{Z}}nidari{\v{c}}}},\
  }\href@noop {} {\bibfield  {journal} {\bibinfo  {journal} {Reviews of Modern
  Physics}\ }\textbf {\bibinfo {volume} {93}},\ \bibinfo {pages} {025003}
  (\bibinfo {year} {2021})}\BibitemShut {NoStop}%
\bibitem [{\citenamefont {Calabrese}\ and\ \citenamefont
  {Cardy}(2007)}]{calabrese2007entanglement}%
  \BibitemOpen
  \bibfield  {author} {\bibinfo {author} {\bibfnamefont {P.}~\bibnamefont
  {Calabrese}}\ and\ \bibinfo {author} {\bibfnamefont {J.}~\bibnamefont
  {Cardy}},\ }\href@noop {} {\bibfield  {journal} {\bibinfo  {journal} {Journal
  of Statistical Mechanics: Theory and Experiment}\ }\textbf {\bibinfo {volume}
  {2007}},\ \bibinfo {pages} {P10004} (\bibinfo {year} {2007})}\BibitemShut
  {NoStop}%
\bibitem [{\citenamefont {Fagotti}\ and\ \citenamefont
  {Calabrese}(2008)}]{fagotti2008evolution}%
  \BibitemOpen
  \bibfield  {author} {\bibinfo {author} {\bibfnamefont {M.}~\bibnamefont
  {Fagotti}}\ and\ \bibinfo {author} {\bibfnamefont {P.}~\bibnamefont
  {Calabrese}},\ }\href@noop {} {\bibfield  {journal} {\bibinfo  {journal}
  {Physical Review A}\ }\textbf {\bibinfo {volume} {78}},\ \bibinfo {pages}
  {010306} (\bibinfo {year} {2008})}\BibitemShut {NoStop}%
\bibitem [{\citenamefont {Alba}\ and\ \citenamefont
  {Calabrese}(2017{\natexlab{b}})}]{alba2017renyi}%
  \BibitemOpen
  \bibfield  {author} {\bibinfo {author} {\bibfnamefont {V.}~\bibnamefont
  {Alba}}\ and\ \bibinfo {author} {\bibfnamefont {P.}~\bibnamefont
  {Calabrese}},\ }\href@noop {} {\bibfield  {journal} {\bibinfo  {journal}
  {Journal of Statistical Mechanics: Theory and Experiment}\ }\textbf {\bibinfo
  {volume} {2017}},\ \bibinfo {pages} {113105} (\bibinfo {year}
  {2017}{\natexlab{b}})}\BibitemShut {NoStop}%
\bibitem [{\citenamefont {Alba}(2018)}]{alba2018entanglement}%
  \BibitemOpen
  \bibfield  {author} {\bibinfo {author} {\bibfnamefont {V.}~\bibnamefont
  {Alba}},\ }\href@noop {} {\bibfield  {journal} {\bibinfo  {journal} {Physical
  Review B}\ }\textbf {\bibinfo {volume} {97}},\ \bibinfo {pages} {245135}
  (\bibinfo {year} {2018})}\BibitemShut {NoStop}%
\bibitem [{\citenamefont {Alba}(2020)}]{alba2020diffusion}%
  \BibitemOpen
  \bibfield  {author} {\bibinfo {author} {\bibfnamefont {V.}~\bibnamefont
  {Alba}},\ }\href@noop {} {\bibfield  {journal} {\bibinfo  {journal} {arXiv
  preprint arXiv:2006.02788}\ } (\bibinfo {year} {2020})}\BibitemShut {NoStop}%
\bibitem [{\citenamefont {Bertini}\ \emph {et~al.}(2016)\citenamefont
  {Bertini}, \citenamefont {Collura}, \citenamefont {De~Nardis},\ and\
  \citenamefont {Fagotti}}]{PhysRevLett.117.207201}%
  \BibitemOpen
  \bibfield  {author} {\bibinfo {author} {\bibfnamefont {B.}~\bibnamefont
  {Bertini}}, \bibinfo {author} {\bibfnamefont {M.}~\bibnamefont {Collura}},
  \bibinfo {author} {\bibfnamefont {J.}~\bibnamefont {De~Nardis}}, \ and\
  \bibinfo {author} {\bibfnamefont {M.}~\bibnamefont {Fagotti}},\ }\href
  {\doibase 10.1103/PhysRevLett.117.207201} {\bibfield  {journal} {\bibinfo
  {journal} {Phys. Rev. Lett.}\ }\textbf {\bibinfo {volume} {117}},\ \bibinfo
  {pages} {207201} (\bibinfo {year} {2016})}\BibitemShut {NoStop}%
\bibitem [{\citenamefont {Castro-Alvaredo}\ \emph {et~al.}(2016)\citenamefont
  {Castro-Alvaredo}, \citenamefont {Doyon},\ and\ \citenamefont
  {Yoshimura}}]{castro2016emergent}%
  \BibitemOpen
  \bibfield  {author} {\bibinfo {author} {\bibfnamefont {O.~A.}\ \bibnamefont
  {Castro-Alvaredo}}, \bibinfo {author} {\bibfnamefont {B.}~\bibnamefont
  {Doyon}}, \ and\ \bibinfo {author} {\bibfnamefont {T.}~\bibnamefont
  {Yoshimura}},\ }\href@noop {} {\bibfield  {journal} {\bibinfo  {journal}
  {Physical Review X}\ }\textbf {\bibinfo {volume} {6}},\ \bibinfo {pages}
  {041065} (\bibinfo {year} {2016})}\BibitemShut {NoStop}%
\bibitem [{\citenamefont {Doyon}(2019)}]{doyon2019lecture}%
  \BibitemOpen
  \bibfield  {author} {\bibinfo {author} {\bibfnamefont {B.}~\bibnamefont
  {Doyon}},\ }\href@noop {} {\bibfield  {journal} {\bibinfo  {journal} {arXiv
  preprint arXiv:1912.08496}\ } (\bibinfo {year} {2019})}\BibitemShut {NoStop}%
\bibitem [{\citenamefont {Ilievski}\ \emph {et~al.}(2018)\citenamefont
  {Ilievski}, \citenamefont {De~Nardis}, \citenamefont {Medenjak},\ and\
  \citenamefont {Prosen}}]{PhysRevLett.121.230602}%
  \BibitemOpen
  \bibfield  {author} {\bibinfo {author} {\bibfnamefont {E.}~\bibnamefont
  {Ilievski}}, \bibinfo {author} {\bibfnamefont {J.}~\bibnamefont {De~Nardis}},
  \bibinfo {author} {\bibfnamefont {M.}~\bibnamefont {Medenjak}}, \ and\
  \bibinfo {author} {\bibfnamefont {T.~c.~v.}\ \bibnamefont {Prosen}},\ }\href
  {\doibase 10.1103/PhysRevLett.121.230602} {\bibfield  {journal} {\bibinfo
  {journal} {Phys. Rev. Lett.}\ }\textbf {\bibinfo {volume} {121}},\ \bibinfo
  {pages} {230602} (\bibinfo {year} {2018})}\BibitemShut {NoStop}%
\bibitem [{\citenamefont {Gopalakrishnan}\ and\ \citenamefont
  {Vasseur}(2019)}]{PhysRevLett.122.127202}%
  \BibitemOpen
  \bibfield  {author} {\bibinfo {author} {\bibfnamefont {S.}~\bibnamefont
  {Gopalakrishnan}}\ and\ \bibinfo {author} {\bibfnamefont {R.}~\bibnamefont
  {Vasseur}},\ }\href {\doibase 10.1103/PhysRevLett.122.127202} {\bibfield
  {journal} {\bibinfo  {journal} {Phys. Rev. Lett.}\ }\textbf {\bibinfo
  {volume} {122}},\ \bibinfo {pages} {127202} (\bibinfo {year}
  {2019})}\BibitemShut {NoStop}%
\bibitem [{\citenamefont {Nardis}()}]{privateCommJacopo}%
  \BibitemOpen
  \bibfield  {author} {\bibinfo {author} {\bibfnamefont {J.~D.}\ \bibnamefont
  {Nardis}},\ }\href@noop {} {\bibinfo  {journal} {Private communication}\
  }\BibitemShut {NoStop}%
\end{thebibliography}%

\end{document}